\documentclass{article}

    \PassOptionsToPackage{numbers,sort&compress}{natbib}

\usepackage[preprint]{neurips_2021}

\usepackage[utf8]{inputenc} 
\usepackage[T1]{fontenc}    
\usepackage{hyperref}       
\usepackage{url}            
\usepackage{booktabs}       
\usepackage{amsfonts}       
\usepackage{nicefrac}       
\usepackage{microtype}      
\usepackage{xcolor}         
\usepackage{notoccite}

\usepackage[pdftex]{graphicx}
\usepackage{algorithm}
\usepackage{algpseudocode}
\usepackage{multirow}
\usepackage{multicol}
\usepackage{listings}
\usepackage{xspace}
\usepackage{subcaption}

\usepackage{amsmath}
\DeclareMathOperator*{\argmax}{arg\,max}

\usepackage{bbm}

\DeclareMathAlphabet{\mymathbb}{U}{BOONDOX-ds}{m}{n}

\newcommand{\elbo}{\mathrm{ELBO}}
\newcommand{\KL}{\mathrm{KL}}
\newcommand{\ourmodel}{NWT\xspace}

\title{\ourmodel: Towards natural audio-to-video generation with representation learning}

\author{%

  Rayhane~Mama \\
  Cash App Labs \\
  Toronto \\
  \texttt{rayhane@squareup.com} \\
  \And
  Marc S.~Tyndel \\
  Cash App Labs \\
  Toronto \\
  \texttt{mtyndel@squareup.com} \\
  \And
  Hashiam~Kadhim \\
  Cash App Labs \\
  Toronto \\
  \texttt{hashiam@squareup.com} \\
  \And
  Cole~Clifford \\
  Cash App Labs \\
  Toronto\\
  \texttt{cclifford@squareup.com} \\
  \And
  Ragavan~Thurairatnam \\
  Cash App Labs \\
  Toronto \\
  \texttt{ragavan@squareup.com} \\

}

\begin{document}

\maketitle

\begin{abstract}
  In this work we introduce \ourmodel, an expressive speech-to-video model. Unlike approaches that use domain-specific intermediate representations such as pose keypoints, \ourmodel learns its own latent representations, with minimal assumptions about the audio and video content. To this end, we propose a novel discrete variational autoencoder with adversarial loss, dVAE-Adv, which learns a new discrete latent representation we call Memcodes. Memcodes are straightforward to implement, require no additional loss terms, are stable to train compared with other approaches, and show evidence of interpretability. To predict on the Memcode space, we use an autoregressive encoder-decoder model conditioned on audio. Additionally, our model can control latent attributes in the generated video that are not annotated in the data. We train \ourmodel on clips from HBO’s Last Week Tonight with John Oliver. \ourmodel consistently scores above other approaches in Mean Opinion Score (MOS) on tests of overall video naturalness, facial naturalness and expressiveness, and lipsync quality. This work sets a strong baseline for generalized audio-to-video synthesis. Samples are available at \href{https://next-week-tonight.github.io/NWT/}{\url{https://next-week-tonight.github.io/NWT/}}. 
  
\end{abstract}

\section{Introduction}

Video generation in deep learning is a challenging domain that lags behind other areas of synthetic media. Recent years have seen incredible generative modeling demonstrations in a variety of tasks such as speech synthesis \cite{hsu2018hierarchical, pngbert}, image generation \cite{ramesh2021zeroshot, salimans2017pixelcnn, chen2020generative}, and natural language text generation \cite{brown2020language}. While there have been interesting recent video synthesis works such as DVD-GAN \cite{clark2019adversarial}, trained on a diverse dataset to produce unconstrained video clips, video generation tasks with a specific generative goal have thus far mostly been accomplished in highly engineered or limited ways. 

There are a number of existing approaches to generating video of a talking person from speech input. To our knowledge, all previous approaches make use of either an engineered intermediate representation such as pose keypoints, or real reference frames as part of the synthesis input, or both. 

Speech2vid in 2017 was the first work to use raw audio data to generate a talking head, modifying frames from a reference image or video \cite{speech2vid}. Another work the same year called ``Synthesizing Obama'' demonstrated substantially better perceptual quality, trained on a single subject \cite{supasorn}. It uses a network to predict lip shape, which is used to synthesize mouth shapes, which are recomposed into video frames using reference frames. Another work, Neural Voice Puppetry aims to be quickly adaptable to multiple subjects with a few minutes of additional video data \cite{Thies2020NeuralVoicePuppetry}. It accomplishes this by using a non-subject specific 3D face model as an intermediate representation, followed by a rendering network that can be quickly tuned to produce subject-specific video. Published in 2020, Wav2Lip puts emphasis on being content-neutral; using reference frames it can produce compelling lipsync results on unseen video \cite{wav2lip_Prajwal_2020}. This also allows it to include hand movement and body movement. Another approach, Speech2Video (not to be confused with the similarly named Speech2vid mentioned above) uses pose as its intermediate representation and a labelled pose dictionary for each subject \cite{liao2020speech2video}.

Engineered intermediate representations have the problem of constraining the output space. The video rendering stage must make many assumptions from face keypoints, a representation with low information content, in order to construct expressive motion. Small differences in facial expression can communicate highly significant expressive distinctions \cite{warsawfacialexpressions}, resulting in important expressive variations being represented with very little change in most intermediate representations of a face. As a result, even if those differences are actually highly correlated with learned features from the audio input, the video rendering stage must accomplish the difficult task of detecting them from a weak signal. 

Models that make use of reference frames, on the other hand, need manual selection of reference content in order for gestures and expression to match the audio. Most models that use reference frames cannot know if, say, an eyebrow raise should not accompany a particular segment of speech. Speech2Video addresses this problem by allowing the model to select from a dictionary of labelled poses. This improves audio-video cohesion but requires extensive data preparation and results in expressive range that can be limited by the scope of the reference dictionary. 

Taking inspiration from recent advances in representation learning \cite{bengio2013representation, oord2018neural}, we took the approach of separating audio-to-video synthesis into two stages. Critically, unlike approaches that use engineered spaces, we employ a fully learned intermediate representation, requiring minimal assumptions about what information is most important for reconstructing video. We train on one subject, John Oliver, but our data include clips spanning years of his show with noticeable variation in camera setup, lighting, clothing, hair, seating position, and age. 

To accomplish this, we introduce the following contributions:

\begin{itemize}
    \item a novel variational discrete autoencoder model, dVAE-Adv, including:
    \begin{itemize}
    \item a new discrete latent structure, Memcodes, whose elements are sampled from a combination of multinomials.
    \item a training loss function for variational autoencoders (VAEs) that avoids averaging tendencies, or ``blur''.
    \end{itemize}
    \item an autoregressive sequence-to-sequence model that generates video from speech using the Memcode video space learned by the dVAE-Adv model.
\end{itemize}

Because \ourmodel makes minimal assumptions about what should be captured in its intermediate representation, it is capable of learning subtle expressive behaviors that might be lost in engineered representations. And because we do not use reference video frames as input, \ourmodel renders full frames from scratch, meaning all gestures and motion are inferred from speech. \ourmodel is also capable of controlling labelled and unlabelled abstract properties of the output. The result\footnote{\url{https://next-week-tonight.github.io/NWT/}} is strikingly natural fully synthesized video.

\section{Method} \label{Method}
Our approach aims to create a generalizable audio-to-video synthesis model that assumes little about the content of either domain. Since autoregressive generative models have shown success in many domains \cite{brown2020language, salimans2017pixelcnn, oord2016wavenet}, we propose training an autoregressive video generator conditioned on each clip's corresponding audio. However, training such a model directly on pixels quickly results in scaling issues that constrain model size and video length. Additionally, autoregressive models trained with likelihood maximization on the pixel domain tend to dedicate a portion of their parameters to learning high frequency details that are perceptually less important than low frequency structure.

To address these two problems, we take inspiration from the representation learning ideas in VQ-VAE \cite{oord2018neural} and split our approach into two independently trained models:

\begin{itemize}
\item \textbf{dVAE-Adv} - A novel discrete variational autoencoder with an adversarial loss that compresses video frames from $256\times 224$ to a $16\times 14$ latent space. \textbf{We call each resulting latent grid element a Memcode.} Despite each Memcode carrying the information of about $768$ elements in the pixel domain, we do not see a significant perceptual degradation in the visual quality of the reconstruction as shown in Section \ref{video_compression_section}.
 
\item \textbf{Prior autogressive model} - An encoder-decoder audio-to-video Memcode prediction model that autoregressively samples from the categorical distribution over the discrete representation. Since this model operates on the Memcode space, memory requirements are low and the network can avoid overspending parameters on high frequency artifacts.
\end{itemize}

Comparable approaches have been used in image \cite{razavi2019generating, ramesh2021zeroshot}, audio \cite{oord2018neural}, and video \cite{walker2021predicting, wu2021godiva} generation.

\begin{figure}
  \centering
  \begin{subfigure}{0.62\textwidth}
    \includegraphics[width=\linewidth]{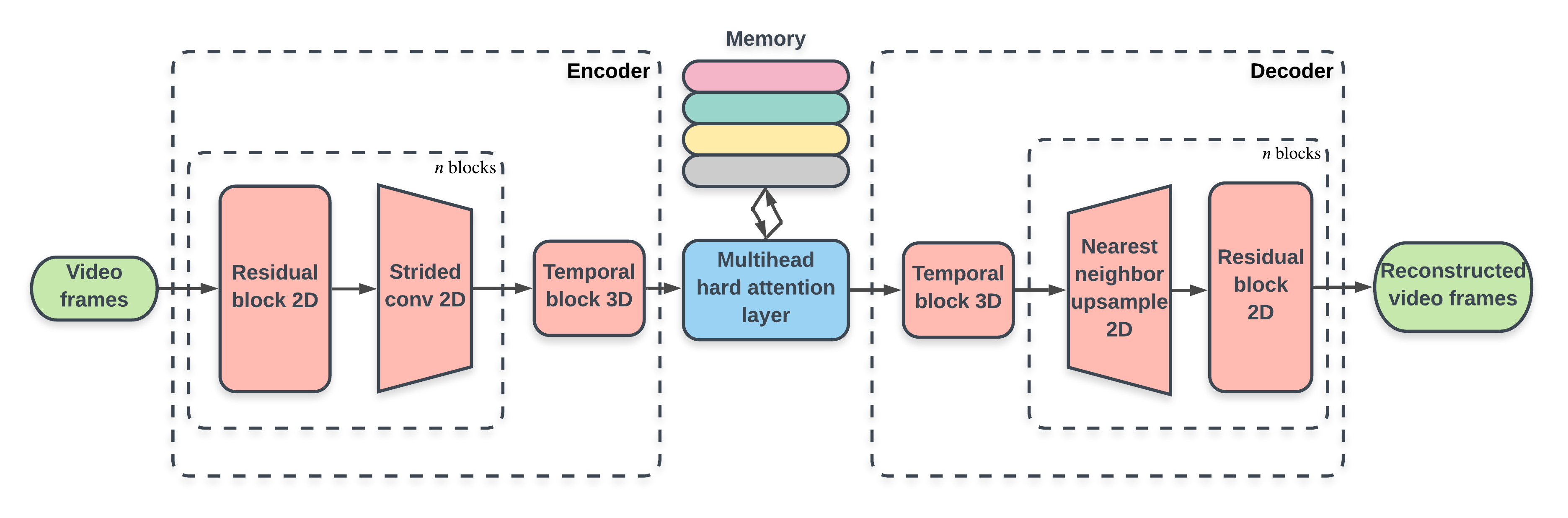}
    \caption{dVAE-Adv}
    \label{model:a}
  \end{subfigure}%
  \hspace*{\fill}   
  \begin{subfigure}{0.36\textwidth}
    \includegraphics[width=\linewidth]{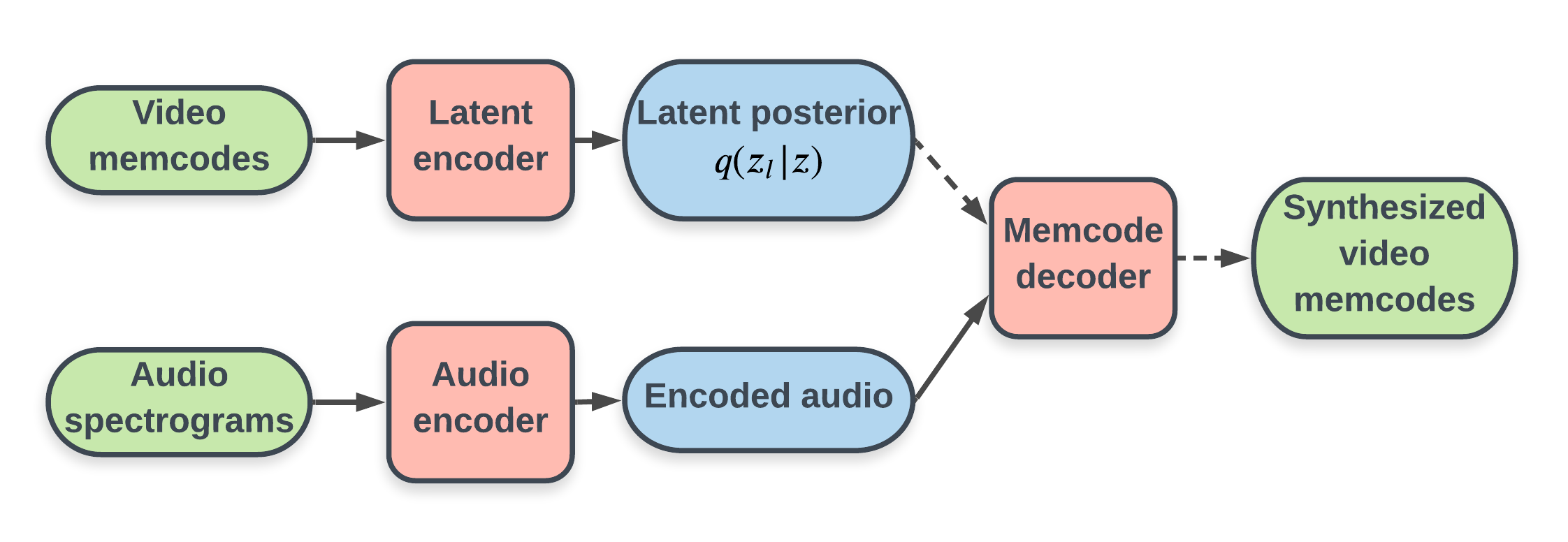}
    \caption{Prior autoregressive model}
    \label{model:b}
  \end{subfigure}
  \caption{(a) dVAE-Adv architecture. The model creates a set of queries using its encoder, which do a lookup from a discrete embedding space (memory). The decoder builds its outputs from only the retrieved vectors. (b) Prior autoregressive model architecture. Dashed arrows denote sampling.}
\end{figure}

\subsection{dVAE-Adv model}
First, we train the dVAE-Adv model with a latent Memcode space of $16\times14$ video frames. This model is trained on the video domain only, ignoring audio.

\subsubsection{Attention as a multinomial discretization function} \label{disc_func}

 Sampling from a multinomial distribution is not differentiable, which prior works have addressed in a number of ways. Some approaches have used a Gumbel-softmax relaxation to gradually push a softmax distribution towards a categorical one as temperature decreases \cite{jang2017categorical, maddison2017concrete, unsupervisedspeechrecognition}. Other discretization schemes rely on a nearest neighbor selection coupled with a straight-through gradient estimator \cite{oord2018neural, razavi2019generating, bengio2013estimating}. While Gumbel-softmax relaxation requires delicate tuning of temperature annealing parameters, nearest neighbor selection requires the addition of both a vector quantization objective and a commitment loss.

We propose a new discretization technique with a gradient heuristic that learns a combination of multinomial distributions without requiring extra loss terms or tuning of temperature parameters.  

We start by using the Gumbel-max trick to sample from a softmax distribution \cite{kool2019stochastic, NIPS2014maddisson}. The softmax is defined by a vector of logits $\chi$, over our model's embedding codebook of size $N$:

\begin{equation} \label{gumbel_max_trick}
\tilde{\alpha} = \argmax_{i \in \{1,\ldots,N\}}\left(S_{\max}\left(\chi + G \right) \right)
\end{equation}
where $G$ is the Gumbel noise and $S_\mathrm{max}$ is the softmax function.

Since $\argmax$ is not differentiable, we apply a straight-through gradient estimation. We find that this approximation works well in practice and is easy to implement in deep learning frameworks. We provide, in Algorithm \ref{diff_argmax}, a simple gradient stopping trick to implement our trainable sampling.

Although we could write (\ref{gumbel_max_trick}) without the softmax function, it is necessary during training to maintain gradient stability. The softmax can be omitted in inference.


In Appendix \ref{gradient_approx}, we justify our design choice to use the Gumbel-max trick over performing natural sampling from the softmax distribution coupled with a straight-through gradient estimation.


We compute the log-probabilities $\chi$ as a content-based similarity measure between the encoder's outputs and the model's latent codebook. We base this similarity metric on the scaled dot-product attention where the queries $Q$ are the projected encoder's outputs, and the keys $K$ and values $V$ are linear projections of the discrete codebook \cite{vaswani2017attention}.

We thus define our entire discretization layer as:

\begin{equation}\mathrm{HardAttn}(Q, K, V) = e_{\tilde{\alpha}} V\end{equation}
\begin{equation} \label{hard_max_equation}
\tilde{\alpha} = \argmax_{i \in \{1,\ldots,N\}}\left(\alpha \right);\;\;\;\; \alpha = S_{\max}\left(\frac{QK^T}{\sqrt{d_K}} + G \right)
\end{equation}

where $e_{\tilde{\alpha}}$ is the standard basis vector in $\mathbb{R}^N$ (one hot encoding of $\tilde{\alpha}$).

Since we want each code to consist of multiple multinomial samples, we use multi-headed attention in our discretization layer \cite{vaswani2017attention}. This yields three benefits over a single softmax distribution:
\begin{itemize}
\item The structure enables $N^h$ permutations per Memcode where $h$ is the number of heads, allowing a large discrete space using a small codebook size.
\item Similar to multivariate Gaussian latents with a diagonal covariance matrix, multi-head multinomial sampling can learn interpretable high level representations of inputs (see \ref{appendix_semantic_representation}).
\item It enables a distance measure between Memcode values, and we find the Hamming distance to be well correlated with loss (see \ref{appendix_categorical_distance}).
\end{itemize}

\subsubsection{Learning} \label{vae_gan_train}
We start by defining the objective of dVAE-Adv as minimizing the negative evidence lower bound (ELBO) of the marginal likelihood of the data \cite{Kingma2014welling, Higgins2017betaVAELB}:

\begin{equation} \label{elbo_loss}
    \mathcal{L}_{\elbo}(x) = \mathop{\mathbb{E}}_{z \sim q_{\phi}(z|x)} \left[- \log p_{\theta}(x|z)\right] + \beta D_{\KL}(q_{\phi}(z|x) || p(z))
\end{equation}
\begin{equation}
    - \log p(x) \leq \mathcal{L}_{\elbo}(x)
\end{equation}

where $q_{\phi}(z|x)$ and $p_{\theta}(x|z)$ denote the variational encoder and decoder respectively. $\beta$ is the KL-annealing loss weight that we find important to tune on video data  \cite{bowman2016generating}, although ELBO is only defined for $\beta = 1$.



In practice, training VAEs to minimize the negative ELBO alone typically yields models that predict averages over high-frequency details of data. This tendency results in blurry output images for image generation tasks \cite{NEURIPS2018introvae, ramesh2021zeroshot}, overly smoothed videos for video generation tasks \cite{yan2021videogpt}, and muffled audio for audio generation tasks \cite{kumar2019melgan}. Autoregressive decoders have been used to mitigate this issue \cite{Chorowski2019arwavenet}, but predicting outputs one element at a time is slow, especially in the video domain. Another approach, inverse autoregressive flows, requires very deep models and an intricate training setup \cite{kingma2017improving}.

To overcome the high frequency averaging problem, we propose training dVAE-Adv with the addition of a Wasserstein adversarial loss term \cite{arjovsky2017wasserstein}. Letting $D_{c}$ be critic $c$ of $C$ from the set of critics $\mathbb{S}_c$, we define the adversarial loss of dVAE-Adv as:

\begin{equation}
    \mathcal{L}_{\mathrm{gan}} = \sum_{c \in \mathbb{S}_C} \mathop{\mathbb{E}}_{z \sim q_{\phi}(z|x)} \left[ - D_c(p_{\theta}(x|z)) \right]
\end{equation}

Additionally, we replace the expected negative log likelihood from (\ref{elbo_loss}) with a reconstruction error expressed inside the critics, in a similar way to \citet{larsen2016autoencoding}:

\begin{equation}
    \mathcal{L}_{\mathrm{recon}} = \sum_{c \in \mathbb{S}_C} \sum_{l \in \mathbb{S}_{L_{c}}} \mathop{\mathbb{E}}_{z \sim q_{\phi}(z|x)} \left[- \log p_{\theta}(D^{l}_{c}(x)|z) \right]
\end{equation}

\begin{algorithm}
   \caption{Differentiable multinomial sampling}
   \label{diff_argmax}
\begin{algorithmic}
   \State {\bfseries Input:} A vector of logits $\chi$, a softmax function $S_{\max}$, a uniform distribution $\mathcal{U}$, a onehot function $\mathrm{onehot}$, an argmax function $\argmax$, a stop gradient function $SG$
   \State {\bfseries Output:} vector of one-hot weights $\tilde{\alpha}$
   \State $u \sim \mathcal{U}(0, 1)$
   \State $G \leftarrow - \log(-\log(u))$
   \State $\alpha \leftarrow S_{\max}(\chi + G)$
   \State $\tilde{\alpha} \leftarrow \mathrm{onehot}(\argmax(\alpha))$
   \State $\tilde{\alpha} \leftarrow \mathrm{SG}(\tilde{\alpha} - \alpha) + \alpha$
\end{algorithmic}
\end{algorithm}

where $D^{l}_{c}(x)$ denotes the hidden representation of the $l$th layer of critic $c$ from the set of layers $\mathbb{S}_{L_{c}}$ of size $L_c$ for critic $c$. This minimizes the feature matching loss between the critics' feature maps for the targets and generator outputs \cite{kumar2019melgan}. It is also similar to \citet{Hou2019}, but does not require a pretrained VGGNet to compute the feature maps \cite{simonyan2015deep}.

We evaluate our reconstruction loss under the Laplace distribution resulting in a simple $L_1$ loss and we assume the prior $p(z)$ to be uniform over the latent categories. We train all of dVAE-Adv's generator model parameters ($\theta$ and $\phi$) with the final loss function:

\begin{equation} \label{generator_loss}
    \mathcal{L}_\mathrm{gen} = \mathcal{L}_{\mathrm{recon}} + \gamma \mathcal{L}_{\mathrm{gan}} + \beta D_\KL \left(q_{\phi}(z|x) || p(z)\right)
\end{equation}

where $\gamma$ is used to weight the contribution of the adversarial loss compared to the reconstruction loss.

Finally, we train our set of critics on the improved Wasserstein GAN loss \cite{gulrajani2017improved}:

\begin{equation} \label{critic_loss}
    \mathcal{L}_{\mathrm{adv}} = \sum_{c \in \mathbb{S}_C} \left( \mathop{\mathbb{E}}_{z \sim q_{\phi}(z|x)} \left[ D_c(p_{\theta}(x|z)) \right]  - \mathop{\mathbb{E}}_{{x} \sim \mathbb{P}_r}[D_c(x)] + \lambda \mathop{\mathbb{E}}_{\hat{x} \sim \mathbb{P}_{\hat{x}}} \left[\left(\left \| \nabla_{\hat{x}}D_c(\hat{x}) \right \|_2 - 1\right)^2\right] \right)
\end{equation}

where $\hat{x}$ is sampled uniformly along straight lines between real samples from data distribution $\mathbb{P}_r$ and generated samples from generator distribution $p_{\theta}(x|z)$. We implicitly define this distribution as $\mathbb{P}_{\hat{x}}$. We present an overview of the training procedure in Appendix \ref{appendix_training_alg}.


Our approach is similar to VAE-GAN \cite{larsen2016autoencoding}, but VAE-GAN does an additional decoding step on $z_p$, sampled from prior $p(z)$. There, the goal is generating realistic video from the uniform prior. This slows down training and is unnecessary for our task since we are not looking to sample from $p(z)$. 

Furthermore, training the decoder on samples from the prior encourages it to generate videos that fool the critics, but not necessarily faithfully reconstruct the input videos \cite{larsen2016autoencoding}. We expect that this would adversely impact lipsync performance in our downstream task and we found the realism versus reconstruction trade-off to be fully controllable by $\gamma$ without the need for an extra decoding step.

\subsubsection{Architecture}
Figure \ref{model:a} depicts an overview of the dVAE-Adv architecture. The encoder starts with spatial downsampling of the frames independently, followed by temporal feature extraction, whose outputs query a multi-headed hard attention block. The attention block outputs become the inputs to the decoder, which is a mirror of the encoder, using 2D nearest neighbour upsampling. dVAE-Adv also includes critic models with different parameters, including 3D models (temporal coherence) and 2D models (independent frame quality). A detailed explanation of the dVAE-Adv architecture can be found in Appendix \ref{appendix_dvaeadv_architecture}.

\subsubsection{Attention as a memory-augmentation}
dVAE-Adv can be viewed as a memory-augmented network \cite{graves2014neural} where the embedding codebook is a trainable external memory table that stores key information about the data. During training, the model extracts and memorizes key data features. During inference, the input data guides the memory lookup from the embedding table. It is possible to frame our work as a special case of \citet{oord2018neural} where the similarity metric is learnable instead of nearest neighbors. Our model is also a special case of \citet{wang2018style, lample2019large} where our memory read is discrete, resulting in our attention heads learning more separable characteristics about the inputs (see Appendix \ref{appendix_semantic_representation}).

\subsection{Prior autoregressive model}
In this section, we train an autoregressive conditional model $p_{\psi}(z|y)$ to generate video Memcodes from audio $y$ (Figure \ref{model:b}). We freeze the dVAE-Adv model's weights in order to train the prior autoregressive model. Throughout this work, we use Mel spectrograms as our audio representation \cite{shen2018natural}, a common approach in speech-related learning tasks, but other audio representations should also work.

\subsubsection{Learning}

To create video Memcode targets for the prior autoregressive model, we sample once from the multinomial modeled by the dVAE-Adv encoder $q_{\phi}(z|x)$ for each clip \footnote{We found sampling from $q_{\phi}(z|x)$ yielded more dynamic results than simply picking the $\argmax$.}. As the number of samples grows, this becomes equivalent to using the softmax distribution probabilities as soft targets.

The prior autoregressive model is trained in a supervised fashion to minimize the KL divergence between $q_{\phi}(z|x)$ and $p_{\psi}(z|y)$. Since all of dVAE-Adv's parameters are frozen during this phase and the latent targets $z$ are only sampled once, minimizing the KL divergence is equivalent to minimizing the cross entropy between $z$ and the prior model's predictions.

\subsubsection{Architecture}
We suggest training prior models with two different autoregressive strategies:

\begin{itemize}
    \item \textbf{Frame level AR (FAR):} Autoregressively predict video frames, one frame at a time. The model makes $16\times14\times h$ predictions each decoding step. This mix of autoregressive and parallel behaviour enables real time inference speed on GPUs.
    \item \textbf{Memcode level AR (MAR):} Autoregressively predict one element of the Memcode video frame at a time. The model makes $h$ predictions on every decoding step. This model can be perceived as a fully AR model from the perspective of predicting one image patch at a time.
\end{itemize}

A third strategy, head level AR (HAR), meaning autoregressively predicting each head of each Memcode (one prediction per step) is computationally slow and we did not explore it at this time. 

All of these models are convolutional \cite{alexnet} encoder-decoder models. Despite their successes in other domains \cite{vaswani2017attention, brown2020language, wu2021godiva, ramesh2021zeroshot}, we hypothesize that transformers would not provide substantial benefit in our case: transformers excel at learning long term relationships but need more memory to do so, which is not ideal under our hardware constraints.

The architecture (Figure \ref{model:b}) consists of three components: an audio encoder that analyses audio spectrogram inputs, a video decoder which autoregressively predicts elements one by one (frames for FAR, Memcodes for MAR), and a variational latent encoder that creates a highly compressed, fixed size, multivariate Gaussian latent style representation $z_l$ of the target Memcodes. The latent encoder enables the model to learn characteristics of the target that are poorly correlated with the audio input. Appendix \ref{appendix_parm_architecture} contains a more detailed explanation of the architecture and latent style representation.

\begin{table}
\caption{Mean Opinion Score (MOS) evaluations with 95\% confidence intervals computed from the t-distribution. \ourmodel $X$ Samp stands for style sampling on $X$ type model, while \ourmodel $X$ COND stands for style copying on $X$ type model (style sampling and copying explained in Section \ref{generation_control_section})}
\label{multirow_mos}
\begin{center}
\begin{tabular}{llll}
\toprule
Model & Overall Naturalness & Face Naturalness & Lipsync Accuracy   \\
\midrule
 Talking Face    & -- & $1.419 \pm 0.052$ & -- \\
 Wav2Lip + GAN    & -- & -- & $2.058 \pm 0.070$ \\
\midrule

 \ourmodel FAR Samp & $2.402 \pm 0.077$ & $3.570 \pm 0.090$ & $3.662 \pm 0.074$\\
 \ourmodel FAR Cond & $2.481 \pm 0.079$ & $3.622 \pm 0.087$ & $3.666 \pm 0.076$\\
 \ourmodel MAR Samp & $\bf{2.835 \pm 0.075}$ & $\bf{3.915 \pm 0.079}$ &  $\bf{3.872 \pm 0.071}$\\
 \ourmodel MAR Cond & $\bf{2.837 \pm 0.078}$ & $\bf{3.960 \pm 0.079}$ & $\bf{3.883 \pm 0.068}$\\
\midrule

 Ground Truth    &  $4.798 \pm 0.026$ & $4.802 \pm 0.027$ & $4.787 \pm 0.023$ \\

\bottomrule

\end{tabular}
\end{center}
\end{table}


\section{Experiments and results} \label{experiments}
\subsection{Dataset}\label{dataset}
Our dataset consists of cleaned video samples of \textit{Last Week Tonight with John Oliver} (LWT). The total dataset consists of 33.4 hours of video, collected from $215$ episodes and split into $16127$ video clips of an average length $7.46$ seconds, interpolated to 30 FPS. See Appendix \ref{dataprep_details_appendix} for more details including the collection and cleaning process.

\subsection{Model details}

Among related previous work in talking human video generation, only LumièreNet \cite{lumiere} and Speech2Video \cite{liao2020speech2video} generate all pixels of the output video without any reference frames as input. However, LumièreNet uses body heatmaps that omit intra-facial information, and Speech2Video requires a large human-labelled dictionary of reference poses mapped to vocabulary, which we do not have for our data. Since these issues prevent us from performing direct comparisons using the same data, we compare our general audio-to-video model with two models for specific tasks: Wav2Lip \cite{wav2lip_Prajwal_2020} and ``Audio-driven Talking Face Video Generation'' \cite{yi2020audiodriven} which we refer to as Talking Face. For fairness, we design separate metrics when comparing each of these models, explained in \ref{eval}.

We train dVAE-Adv with Adam optimizer \cite{adamoptimizer} for $150,000$ steps with a batch size of 16 videos each containing 30 frames distributed across 16 NVIDIA A100 40GB GPUs. We found that the model is able to generate arbitrary video length without quality decay during inference. More details and the dVAE-Adv model parameters are in Appendix \ref{dvaeadv_parameters_appndex}

We train our prior autoregressive models with Adam optimizer for $250,000$ steps with a batch size of 64 distributed across 8 NVIDIA A100 GPUs. Each sample is limited to $32$ frames during training. For a more detailed description including model parameters, see Appendix \ref{parm_parameters_appendix}. 

For the Talking Face model, we fine-tune the memory-augmented GAN version of Talking Face as instructed in its documentation on our LWT dataset. We use training and inference parameters as recommended in the authors' codebase. To avoid discrepancy between the training and inference distributions, we ensure the reference videos and test audio are sourced from the same episode.

For the Wav2Lip model, we tried both the zero-shot approach using pre-trained models and training from scratch on our dataset, and found the former, specifically the pre-trained model with an adversary, was better. Similarly to Talking Face, we select a random training sample from the same episode as the test audio input to use as the reference input.

\subsection{Evaluation} \label{eval}
We conducted three separate subjective mean opinion score (MOS) experiments. In the first experiment, participants were instructed to rate the overall naturalness of the generated clip; in the second, to specifically rate the face's naturalness and expressiveness; and in the third, to rate the naturalness and accuracy of the lipsync. The bottom half of videos were removed for the latter two so that participants focus on the face and mouth. Table \ref{multirow_mos} shows comparisons between our models and relevant baselines. In the first experiment, we compare \ourmodel with ground truth videos. In the second, we compare \ourmodel with Talking Face \cite{yi2020audiodriven} and the ground truth videos. In the last experiment, we compare \ourmodel with Wav2Lip + GAN \cite{wav2lip_Prajwal_2020}, as well as ground truth videos. Details about our MOS setup are in Appendix \ref{mos_appendix}.

As seen in Table \ref{multirow_mos}, our end-to-end models, which are not explicitly trained for the lipsync task, outperform the Wav2Lip + GAN lipsync model. By manually inspecting generated samples from both models, we find that \ourmodel generates more natural and smooth mouth movements. Wav2Lip + GAN's quality can drop when there is greater movement of the subject's head in the reference video, and from time to time the mouth detection fails resulting in bad outputs. \ourmodel does not suffer from these issues since the full video is generated from scratch.

NWT also significantly outperforms the Talking Face model, which struggles to correctly render a natural face on volatile reference heads. In contrast, our method generates a large variety of facial expressions that pair naturally with the input audio. We demonstrate in Samples Section \href{https://next-week-tonight.github.io/NWT//#style_sampling}{S1} that NWT generates facial expressions and body language that are tightly connected to tone of speech.

\hypertarget{limitation_1}{The} overall naturalness MOS experiment shows that, despite our model being a strong baseline for natural human audio-to-video generation, there is still a sizable gap between real videos and our model. This is likely due to \ourmodel's shortcomings in rendering Oliver's hands and gestures. Our interpretation of this limitation is that Oliver has a wide variety of gestures that have relatively weaker correlations with speech audio features (compared to lip position and head movement), and the data size is too small to learn this well.  We expect our model would perform better in this regard if trained on less emotive and dynamic subjects than John Oliver.

Finally, MAR models perform better than FAR models in terms of MOS, which is likely due to the additional stability that MAR models have over FAR. We show quality comparisons of generated videos from these models in samples section \href{https://next-week-tonight.github.io/NWT//#models_comparison}{S4} and we show the training and inference speeds of each model in Appendix \ref{far_vs_mar}.

\begin{table}
\caption{Compression rate comparison of different video compression systems on our test data using structural similarity index measure (SSIM) \cite{ssim}. The Mem16 model has a latent space of $16 \times 14$ and the Mem8 model has a latent space of $8 \times 7$. Each Memcode contains 24 bits of information. }
\label{compression_table}
\begin{center}
\begin{tabular}{llll}
\toprule
Model & Data size (GB) & Relative compression to h264 & Frame SSIM \\
\midrule
 h264    & $0.289$ & $1\times$ & $0.9452$ \\
\midrule

dVAE-Adv Mem16 & $0.223$ & $1.30\times$ & $0.9499$ \\
dVAE-Adv Mem16 + RLE & $0.146$ & $1.98\times$ & $\bf{0.9499}$ \\
dVAE-Adv Mem8  & $0.112$ & $2.58\times$ & $0.9412$ \\
dVAE-Adv Mem8 + RLE & $\bf{0.072}$ & $\bf{4.01\times}$ & $0.9412$ \\

\midrule

 Uncompressed    &  $57.83$ & -- & $1.0$\\

\bottomrule

\end{tabular}
\end{center}
\end{table}

\subsection{Episode control}
As the data contain 215 episodes varying in camera setup, lighting, clothing, hair, position, and age, we propose a hierarchical episode embedding for both dVAE-Adv and the prior autoregressive model, similar to speaker embeddings in multi-speaker text-to-speech models \cite{skerryryan2018endtoend, oord2016wavenet} (Appendix \ref{episode_embedding_appendix}).

We show in Samples Section \href{https://next-week-tonight.github.io/NWT//#episode_control}{S2} that we can generate video clips from different episodes given the same audio input. In these examples, the episode embedding of dVAE-Adv alone mainly changes global color filters, while the combination of episode embeddings from both models perform a complete episode translation. We hypothesize this behaviour would generalize to multi-actor datasets.

\subsection{Style control}\label{generation_control_section}
In this section, we examine the properties of the unsupervisedly learned style latent representations $z_l$, and demonstrate how they control features that weakly depend on audio (Samples Section \href{https://next-week-tonight.github.io/NWT//#style_control}{S3}).

To understand the effect of the style latent vectors, we randomly sampled an initial vector from the prior $p(z_l)$. We then manually change one variate from the latent vector and analyze the results. We show that some variates control human-interpretable features, such as camera angles or the presence of side overlay boxes from the show.

We further demonstrate that we can provide an optional video reference input $x_r$ to the model to guide the style of the output video. We sample $z_l$ from the posterior $q(z_l|z_r)$ where $z_r$ is sampled from the posterior $q_\phi(z_r|x_r)$ (see \ref{appendix_parm_architecture}). We show that, for an audio input that is different from the audio of the reference $x_r$, the model copies relevant abstract traits from $x_r$ in the generated sample.



\subsection{Video Compression}\label{video_compression_section}
A byproduct of this work is neural video compression. Memcodes act as an efficient representation of video data that lies within the distribution of the training data. In Table \ref{compression_table} we compare dVAE-Adv models with two downsampling rates to the more general industry standard compression algorithm h264 \cite{h264} applied on our test set. We find that dVAE-Adv models can achieve higher compression rates than h264 for comparable reconstruction quality measured by mean frame SSIM \cite{ssim}.

With $\argmax$ sampling, we find that most Memcode heads tend not to change between adjacent frames. We can therefore use a lossless run length encoding (RLE) on the latent video representations. dVAE-Adv + RLE models achieve $2\times$ and $4\times$ more compression relative to h264 respectively for $16\times14$ and $8\times7$ latent space dimensions. Both models exhibit comparable SSIM metrics. We show perceptual comparisons of these compression methods in section \href{https://next-week-tonight.github.io/NWT//#video_compression_and_reconstruction}{S5}.



\section{Applications and broader impact} \label{applications}

There are countless exciting applications to flexible video synthesis. It could provide ways to create hyper-realistic show hosts who can be localised to different languages. It presents a new approach to generating dynamic human-like characters in video games. If this approach can be extended to perform a language translation task, it provides a path towards automated sign language interpreting. 

However, like many emerging technologies, there are potential nefarious uses that warrant serious attention. Synthetic media can enable the creation of content deliberately designed to mislead or deceive. Risk mitigation work has mostly been in synthetic media detection \cite{rossler2019faceforensics, bonettini2020video, agarwal2020detecting}. While NWT is more general than face swapping or lipsync techniques, we believe similar detection approaches can prove effective and that it is important for research to take place while these technologies are new.

\section{Future work}

A clear subsequent direction for \ourmodel is training larger models on much bigger datasets consisting of many subjects. The ability of our model to handle the diverse varieties of individual subjects suggests this flexibility is likely possible. Another idea to explore is replacing the episode embedding with a cloning component, enabling the model to mimic a particular episode given a frame or partial image. This could further be extended to models that can imitate speakers or contexts unseen during training.

It would also be interesting to apply our techniques to non-video domains, such as audio generation or image generation tasks. We would also like to explore disentanglement between Memcode heads, or ideas such as hierarchical Memcodes.

\section{Conclusion}

We describe \ourmodel, a new approach for synthesizing realistic, expressive video from audio. We introduced a new discrete autoencoder, dVAE-Adv, highlighting the multi-headed structure of its latent space which we term Memcodes. We showed that this approach, which avoids implicit bias regarding the contents of video or audio in its architecture, significantly outperforms previous specialized approaches to lipsync and face generation tasks in subjective evaluation tests, while also clustering many latent features without supervision. We evaluated our model on several generation control tasks, demonstrating that we can vary episodes and recording setup independently from speech. Our model would also be easy to link to existing text-to-speech models, which would enable end-to-end text-to-video generation.

Our work sets a new standard for expressiveness in talking human video synthesis and shows potential for more general applications in the future.



\begin{ack}
The authors would like to thank Alex Krizhevsky for his mentorship and insightful discussions. The authors also thank Alyssa Kuhnert, Aydin Polat, Joe Palermo, Pippin Lee, Stephen Piron, and Vince Wong for feedback and support. 


\end{ack}
\newpage
\bibliography{bibliography}
\bibliographystyle{icml2020}


\newpage
\appendix

\begin{figure}
  \centering
  \begin{subfigure}{0.63\textwidth}
    \includegraphics[width=\linewidth]{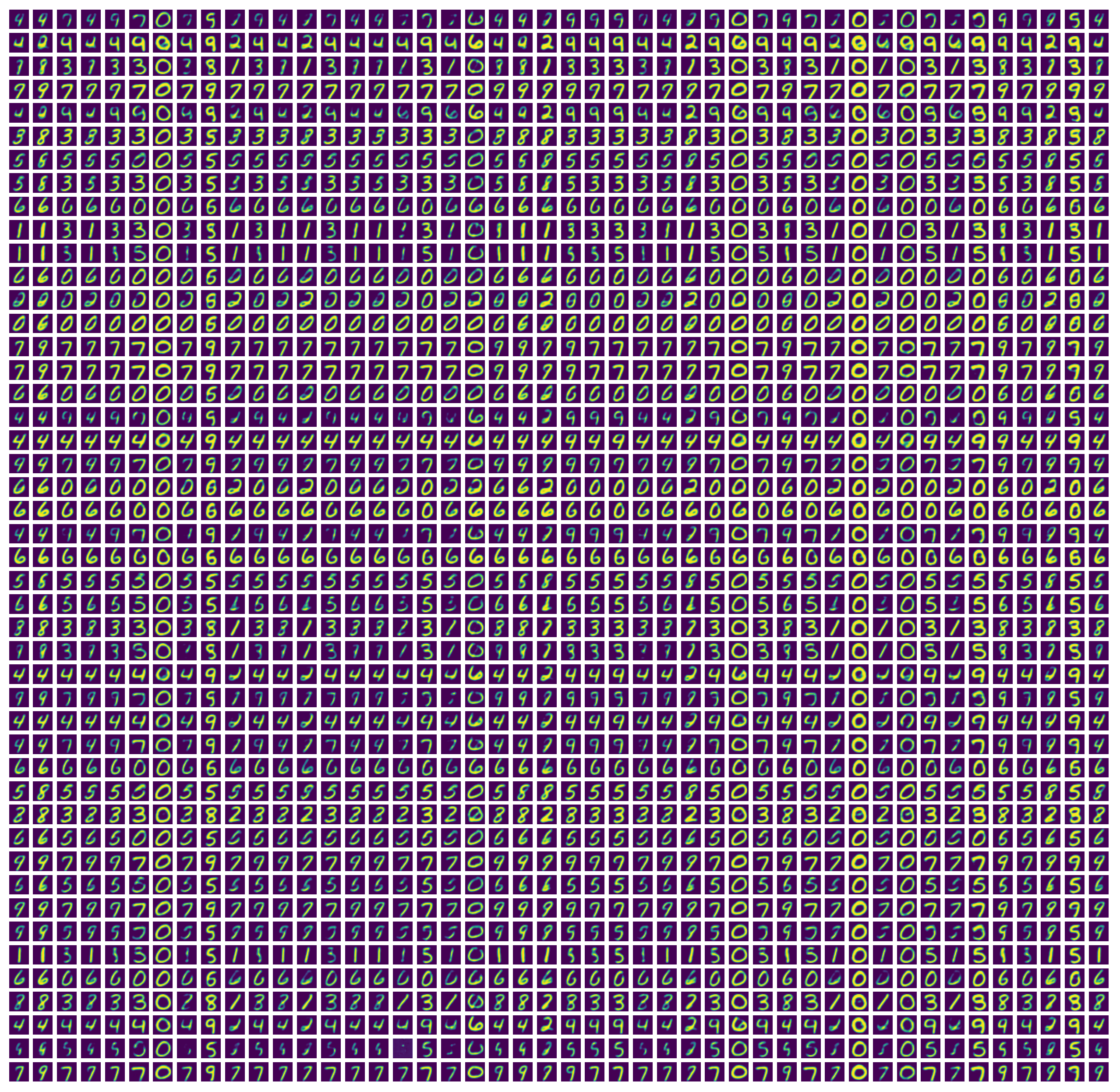}
    \caption{Total possible generations}
    \label{latent_occupancy:a}
  \end{subfigure}%
  \hspace*{\fill}   
  \begin{subfigure}{0.36\textwidth}
    \includegraphics[width=\linewidth]{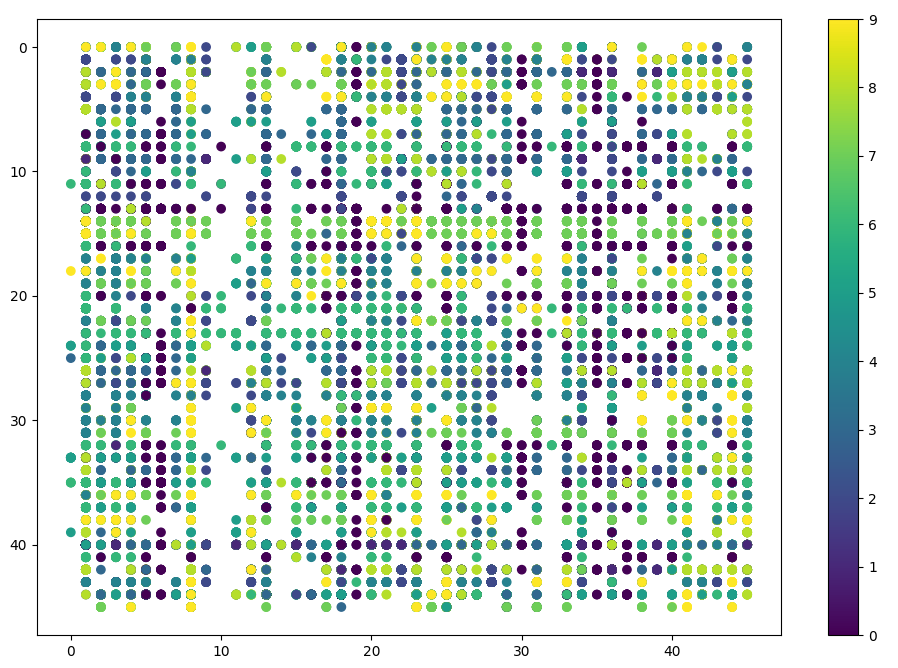}
    \caption{Latent space occupancy}
    \label{latent_occupancy:b}
  \end{subfigure}
  \caption{In both subfigures, the $x$-axis defines the codebook values of head $1$ and the $y$-axis defines the codebook values of head $2$. (a) shows all possible samples the MNIST dVAE-Adv can generate. Since the total latent space is much smaller than the dataset size, the model has to encode minimal information about the images along each head. These features can be related to the digit id or other specific style related to it such as rotation or boldness. (b) shows the Memcode combinations allocated for all MNIST images (with overlap). We find that the MNIST dVAE-Adv occupies most of the Memcode space almost uniformly. The digit color code also shows that some categories, both on heads $1$ and $2$, learn strong biases about the digit id. Some columns only encode the digit $0$ for example.}
\end{figure}

\section{Exploratory Memcode analysis}\label{Memcodes_appendix}
In this appendix, we suggest simple and interpretable experiments on the dVAE-Adv Memcode discretization space to gain insight into what it learns. We conduct these experiments on the MNIST dataset \cite{lecun-mnisthandwrittendigit-2010} since the images are small and we can make a latent space of a single Memcode. Additionally, it is easier to interpret observations on images than videos. 

\subsection{Interpretable representation} \label{appendix_semantic_representation}
We propose to study the behavior of the latent multi-headed discrete space. It is trivial to notice that a single-headed discrete space lacks internal structure per Memcode, so we will use two or more heads for all experiments. 

We start by training a 2-headed dVAE-Adv to reconstruct MNIST with a codebook size of 46. The encoder-decoder parameters are selected to downsample the MNIST images to $1 \times 1$. This is done to simplify analysis. 

We show images generated by setting all possible $46^2$ permutations of the Memcode space in a converged model in Figure \ref{latent_occupancy:a}. We observe that the model learns interpretable representations that can be split by multiple factors such as digit id, rotation degree or boldness. These representations however are not entirely disjoint from each other. A perfectly disentangled representation would have each row (or column) of Figure \ref{latent_occupancy:a} have one and only one digit id, or a distinguishable stylistic element such as boldness, for example. We leave the exploration of ways to train Memcode spaces with fully disentangled heads for future work.

Figure \ref{latent_occupancy:b} shows the latent space occupancy with the MNIST training input data, color coded by digit id. We generate the plot by feeding inputs from the dataset to the dVAE-Adv encoder, then taking the $\argmax$ over the encoder's logits. We find that dVAE-Adv used $83\%$ of the available permutation space. 

\begin{figure}
  \centering
  \begin{subfigure}{0.52\textwidth}
    \includegraphics[width=\linewidth]{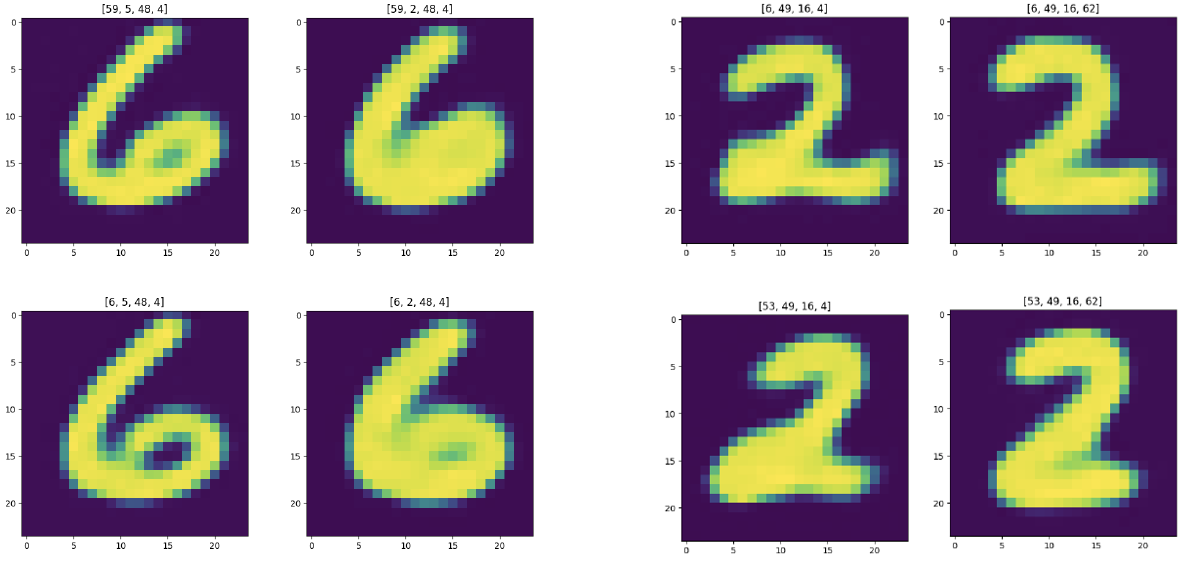}
    \caption{MNIST observations}
    \label{pairwise_comparison:a}
  \end{subfigure}%
  \hspace*{\fill}   
  \begin{subfigure}{0.44\textwidth}
    \includegraphics[width=\linewidth]{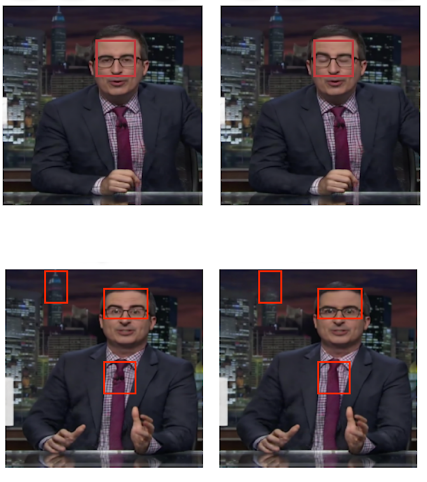}
    \caption{LWT observations}
    \label{pairwise_comparison:b}
  \end{subfigure}%
  \caption{We show that the model learns semantic representations of data in parts of the Memcode space. (a) shows the result of manually changing one head of the Memcode. The MNIST models learn heads responsible for style independently from the digit. (b) shows the same observation on the $16\times 14$ latent space of LWT. Since the latent space doesn't only contain one Memcode, manual modifications here affect only a part of the output frame. We modify either one or a few neighbor Memcodes to apply modifications of different sizes.}
  \label{pairwise_comparison}
\end{figure}

For completeness, we propose encoding MNIST input images, then modifying only one head of the latent Memcode vector to observe its effect on the output. We conduct this on a dVAE-Adv with $4$ heads to gain more dimensions of control. In Figure \ref{pairwise_comparison:a}, we show that we can indeed control the writing style while preserving the digit id by modifying only a part of the Memcode permutation.

We also show examples of an analogous learned structure in LWT data. We take a single frame from the dataset and encode it to the Memcode space. Then we manually change the category of only one head of one or a few of the Memcodes. This results in local interpretable modifications to the input frame (Figure \ref{pairwise_comparison:b}).

\begin{figure}
  \centering
  \begin{subfigure}{0.31\textwidth}
    \includegraphics[width=\linewidth]{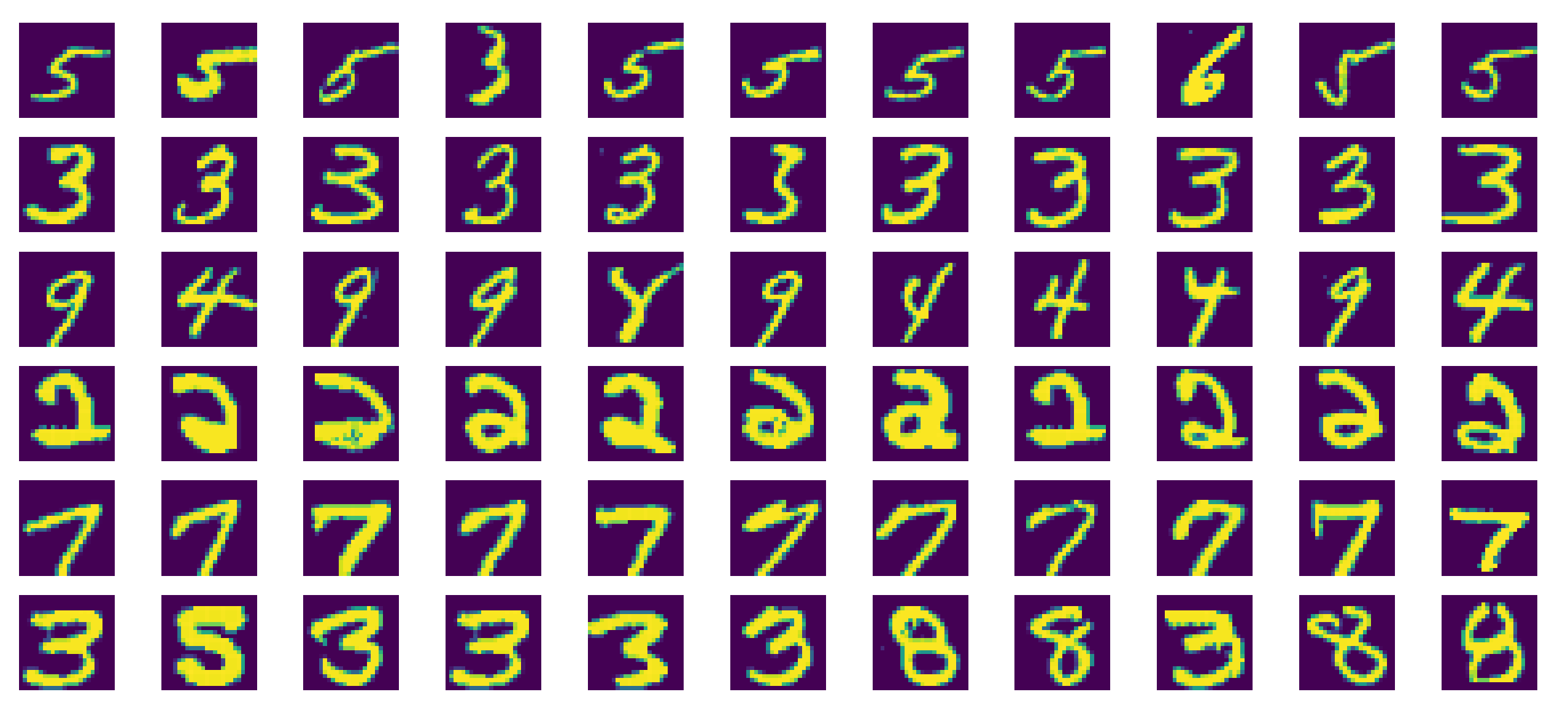}
    \caption{dVAE model}
    \label{pairwise_distance:a}
  \end{subfigure}%
  \hspace*{\fill}   
  \begin{subfigure}{0.31\textwidth}
    \includegraphics[width=\linewidth]{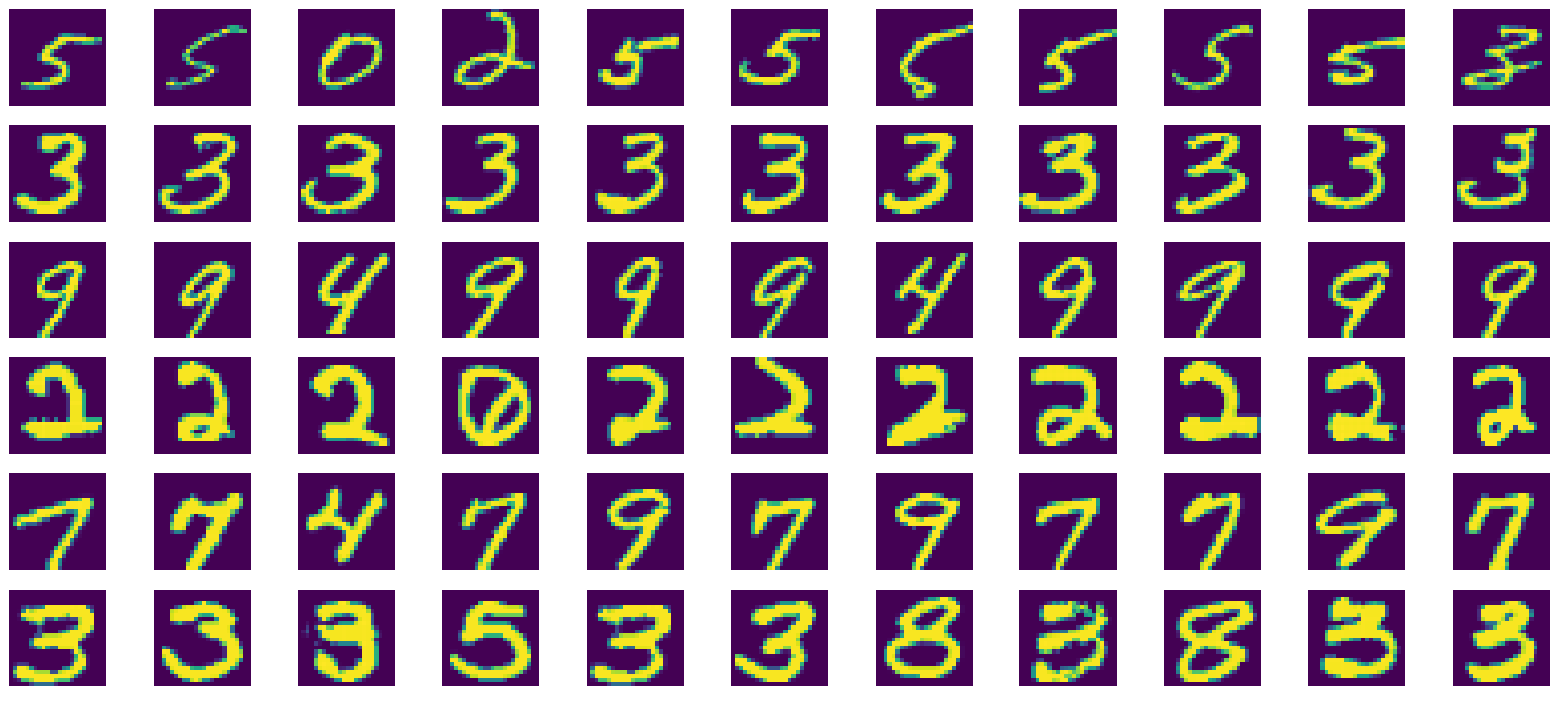}
    \caption{dVAE-Adv model}
    \label{pairwise_distance:b}
  \end{subfigure}%
  \hspace*{\fill}   
  \begin{subfigure}{0.31\textwidth}
    \includegraphics[width=\linewidth]{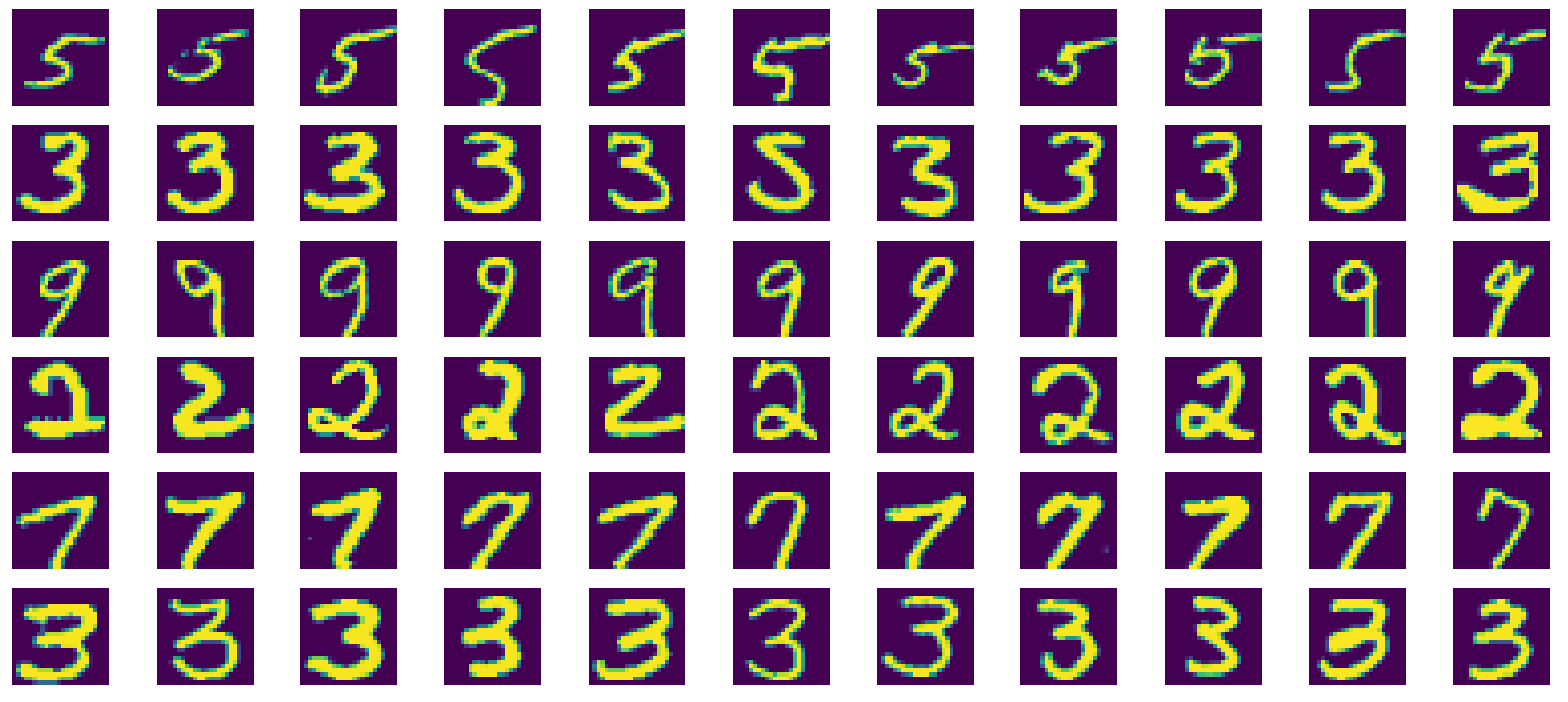}
    \caption{dVAE-Class model}
    \label{pairwise_distance:c}
  \end{subfigure}
  \caption{Image retrieval of 10 nearest neighbors by Hamming distance. In all subfigures, the left column shows the query images fed to the encoder, then the 10 nearest neighbors picked by Hamming distance on the latent Memcode space.}
  \label{pairwise_distance}
\end{figure}

\subsection{Categorical distance} \label{appendix_categorical_distance}

While single-headed discrete VAEs have proved useful in representation learning for generative downstream tasks, such representations are very weak on data retrieval tasks due to lack of distance in the latent space. However, we show in this section how multi-headed discrete VAEs overcome that problem by implicitly learning a distance that is highly correlated to the loss function used to train the VAE.

We define the distance computed between two Memcodes $z_1$ and $z_2$ as the Hamming distance:

\begin{equation}
    \mathcal{L}_\mathrm{ham} = \frac{1}{h} \sum_{i = 0}^{h-1} \left(  {z_1}_i \ne {z_2}_i  \right)
\end{equation}

Two Memcodes are considered closer if they have more heads in common. This does not assume that all data samples with similar features share the same heads, but that samples which do share heads should have definable characteristics in common in the pixel space. The same distance was also used in content-based image retrieval with binary autoencoders \cite{conf/esann/KrizhevskyH11}.

Since the Hamming distance's resolution depends on the number of heads, we train dVAE-Adv on MNIST with a latent space of $1\times 1$, a codebook of size $N=2$ and $11$ heads (total number of permutations $2^{11}$).

To build an intuition around what the Hamming distance represents in practice, we propose feeding a set of inputs to dVAE-Adv's encoder and doing $\argmax$ sampling on its probabilities for every image. We then compute $\mathcal{L}_\mathrm{ham}$ against every other sample in the data and we plot the nearest neighbors by Hamming distance in ascending order.

Figure \ref{pairwise_distance:a} shows the results of nearest neighbors selection on our models trained with the ELBO loss from (\ref{elbo_loss}). We notice that the model clusters samples together based on image similarity on the pixel space. i.e, digits that occupy the same regions in the pixel domain tend to be clustered together in the Memcode domain. This may be due to images having similar degrees of rotation, similar boldness or even similar digits. To quantify the extent to which this metric groups digits together, we compute the zero-shot Hamming distance nearest neighbor classification and get a $75.3\%$ accuracy on the test set.

We then train the same model using the generator loss from (\ref{generator_loss}) where $\mathcal{L}_\mathrm{recon}$ is computed across all but the output layer of the critic. We show the results in Figure \ref{pairwise_distance:b}. We observe that this model clusters the images less by their similarity on the pixel domain, and more by their likelihood to share features. i.e, the generator clusters images by their similarity on the critic's feature maps. Since digits share features between each other, we find that this model's clustering still does not correlate with the digits ids perfectly, which is confirmed by a $76.9\%$ accuracy by zero-shot Hamming distance nearest neighbor classification.

Finally, we propose modifying the critic from dVAE-Adv to also play the role of a digit classifier. We call this full model dVAE-Class. To do so, we add a second output layer to the critic and add a digit classification cross entropy to (\ref{critic_loss}). We note that both critic output layers use the same activations from the penultimate layer as input. We also only optimize $\mathcal{L}_\mathrm{recon}$ on the penultimate layer's activations. We assume that a digit classifier discards sample specific information and only keeps digit id relevant information in the penultimate layer. We show the results of this model in Figure \ref{pairwise_distance:c}. 

In contrast to previous experiments, dVAE-Class clearly clusters the images by digit id and much less by other sample specific features. This is also reflected in a $94.7\%$ accuracy by zero-shot Hamming distance nearest neighbor classification, compared to $97.3\%$ accuracy by the critic classifier. We achieve very similar results by training the generator on the penultimate layer's activations of a pre-trained MNIST classifier, without use of a critic.

These observations encourage us to use a multi-headed categorical distribution over the latent space on generation tasks, since wrongly predicting one of the heads should still generate an output that is close to the target, where the distance depends on the dVAE loss function. We believe that this distance property could also be useful in other clustering tasks where one can design the loss function to create specific data clusters.

\newpage
\section{Model architectures}\label{appendix_architectures}
\subsection{dVAE-Adv}\label{appendix_dvaeadv_architecture}

A variational autoencoder trained to compress and reconstruct videos.

\subsubsection{Encoder}

The dVAE-Adv encoder starts by spatially downsampling the input video on each frame independently using 2D residual blocks \cite{resnet} followed by a strided convolution \cite{bai2018empirical}. If the number of filters changes in the residual block, we use a shortcut $1\times1$ conv in the residual branch. After extracting the spatial feature maps for each frame, we compute the temporal feature maps across frames using a temporal block consisting of a stack of 3D convolutions. We found the temporal block necessary to remove flicker and prevent jitteriness. A large temporal receptive field was not necessary in our experiments. Note that we do not compress the video clip along the time dimension.

\subsubsection{Hard attention discretization layer}

The encoder outputs (temporal feature maps) are used as the query sequence $Q$ for the multi-head hard attention block. The keys $K$ and values $V$ sequences are separate linear projections of an embedding space $e \in{\mathbb{R}^{N \times D}}$ where $D$ is the dimensionality of the embedding codebook. The multi-head hard attention block looks up the embedding table $h$ times (the number of attention heads) for each Memcode. The attention is computed independently between each element of the queries sequence $Q$, and is computed against all rows of the embedding space $e$. 

\subsubsection{Decoder}

The output of the multi-head hard attention is used as the decoder input. The decoder is a mirrored version of the encoder. We first compute the temporal feature maps using a similar architecture to the encoder's temporal block. To output in the original video resolution, we use a 2D nearest neighbor upsampling layer \cite{odena2016deconvolution} followed by residual blocks that operate on each frame independently. The reconstructed spatial feature maps are then passed to a $\tanh$ output layer for the final pixel values.

\subsubsection{Critics}

For the critics, we design each model as a Markovian window-based adversary (analogous to image patches \cite{kumar2019melgan, isola2018imagetoimage}) consisting of a sequence of strided convolutional layers. We ensure that critics have different strides and receptive fields to capture multi-scale information about the videos. We also found it beneficial to make some of the critics operate only spatially to improve independent frame quality. In all our experiments, we compute $\mathcal{L}_\mathrm{recon}$ on all but the output layers of the critics.


\subsubsection{Gradient approximation} \label{gradient_approx}

While the Memcode discretization using Gumbel-max trick as presented in (\ref{gumbel_max_trick}) is equivalent to a natural sampling from the multinomial distribution ($z \sim S_\mathrm{max}(\chi)$), the gradient flow during back propagation is almost always different:

\begin{equation}
    \frac{\partial S_\mathrm{max}(\chi)}{\partial \phi} \ne \frac{\partial S_\mathrm{max}(\chi + G)}{\partial \phi}
\end{equation}

since

\begin{equation}
    \frac{\partial S_\mathrm{max}(\chi)}{\partial \phi} = \frac{\partial S_\mathrm{max}(\chi)}{\partial \chi} \frac{\partial \chi}{\partial \phi}
\end{equation}

and 

\begin{equation}
    \frac{\partial S_\mathrm{max}(\chi + G)}{\partial \phi} = \frac{\partial S_\mathrm{max}(\chi + G)}{\partial (\chi + G)} \underbrace{\frac{\partial (\chi + G)}{\partial \chi}}_{=1} \frac{\partial \chi}{\partial \phi} = \frac{\partial S_\mathrm{max}(\chi + G)}{\partial (\chi + G)} \frac{\partial \chi}{\partial \phi}
\end{equation}

where $\phi$ are the dVAE-Adv encoder parameters and the Gumbel noise $G$ is almost always different from the zero-vector $\mymathbb{0}$.

We expect the model to be more stable to train when the gradient approximation is more representative of the forward pass. From that perspective, a gradient approximation that maximizes the likelihood of a latent $z$ being sampled from the softmax distribution is more stable to train. 

\begin{figure}
  \centering
  \includegraphics[width=\linewidth]{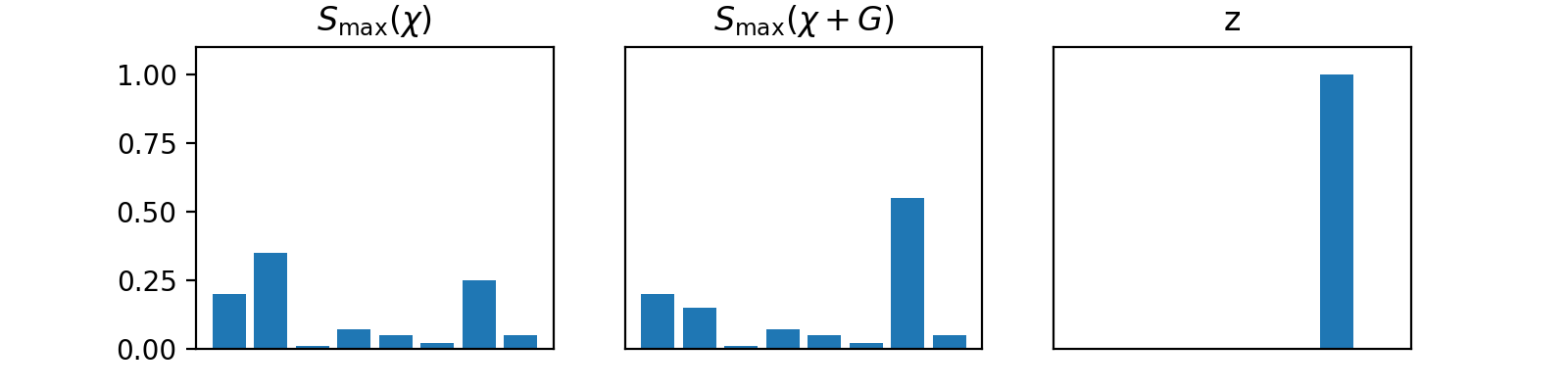}
  \caption{(Left) Softmax distribution over logits $\chi$. (center) The same softmax distribution after the addition of gumbel noise $G$. (right) Sampled $z$ by either multinomial sampling from $z \sim S_\mathrm{max}(\chi)$ or $\argmax(S_\mathrm{max}(\chi + G))$. The sample $z$ is more likely to be sampled from $S_\mathrm{max}(\chi + G)$ than from $S_\mathrm{max}(\chi)$.}
  \label{softmax_figure}
\end{figure}

In categorical distributions, the $\argmax$ sampling guarantees maximum likelihood, which makes the Gumbel-trick all the more interesting since the straight-through gradient estimation only bypasses the $\argmax$ function, instead of the whole sampling operation. This ensures that our gradients, which depends on the Gumbel noise, are more representative of the forward pass (Figure \ref{softmax_figure}).

Empirically, we found that models trained with the Gumbel-max trick explained in (\ref{gumbel_max_trick}) are more stable and converge to slightly better optimums in comparison to sampling from the softmax distribution. The former models are also less prone to gradient explosions that cause the latent space to collapse, even when trained with small batch sizes. 



\subsection{Prior autoregressive model}\label{appendix_parm_architecture}

A prior autoregressive model which predicts video Memcodes from audio (Figure \ref{model:b}). 

\subsubsection{Audio encoder}

The audio encoder uses a sequence of convolutional layers to analyze the audio spectrogram inputs and extract useful features for the video generation task. These feature maps are then upsampled spatially using nearest neighbor upsampling and spatially connected layers. We found this upsampling strategy to be more stable than simply tiling the audio feature maps over the spatial dimensions.

Spatially connected layers are similar to 2D locally connected layers in space dimensions, but share weights on the time dimension. In contrast to simple 3D convolutions, spatially connected layers benefit from space location awareness. In practice, it is very easy to implement this layer as a patch extraction convolution with a non-trainable eye matrix kernel followed by a matrix multiplication. Such an implementation makes this layer almost as fast as a convolutions while having location aware parameters. Empirically, spatially connected layers yielded more dynamic yet more stable results than a simple tile or a convolution upsample.

\subsubsection{Style encoder}

Similarly to \citet{hsu2018hierarchical}, we found it useful to include a variational latent encoder $q(z_l|z)$ that creates a compressed, fixed size, multivariate Gaussian latent representation of the target Memcodes. The style encoder uses a stack of convolutions to extract useful style feature maps from the target Memcodes. To create a fixed size vector from variable length videos, we use a global average pooling layer to aggregate the convolution feature maps. Including this variational encoder enables the model to unsupervisedly learn a latent representation $z_l$ encoding global characteristics of the output videos that do not correlate well with the audio input. We upsample $z_l$ using a similar approach described above for upsampling audio feature maps. 

\begin{table}
\caption{Parameter count for FAR and MAR models, along training and inference speeds. training speed is calculated with a batch size of 64 and a maximum number of frames of 32 per sample spread across 8 GPUs with data parallelization. Inference speed is computed for single sample inference.}
\label{model_speeds}
\begin{center}
\begin{tabular}{lp{3.2cm}p{2cm}p{2.25cm}}
\toprule
Model & \# trainable parameters (in millions) & training speed (updates/sec) & inference speed (frames/sec)   \\
\midrule
 \ourmodel FAR & $383.484$ & $1.39$ & $38.9$ \\
 \ourmodel MAR & $216.327$ & $0.81$ & $0.77$ \\

\bottomrule

\end{tabular}
\end{center}
\end{table}

During training, $z_l$ is sampled from the posterior $q(z_l|z)$, then upsampled and added along with the audio feature maps to the video Memcode decoder. We also minimize the KL divergence between the posterior $q(z_l|z)$ and a standard normal Gaussian with a diagonal covariance matrix $p(z_l) = \mathcal{N}(0, \mathcal{I})$.

During inference, the variational latent encoder enables us to control inference in two ways:
 \begin{itemize}
     \item \textbf{Style copy:} We sample $z_l$ from the posterior of an encoded reference video. We show in Section \ref{generation_control_section}, that the model successfully copies video features that are not directly correlated with the audio.
     \item \textbf{Style sampling:} We randomly sample $z_l$ from the prior distribution $p(z_l)$ and generate different looking videos for the same audio (see \ref{generation_control_section}). We also show the effect of each Gaussian variate independently of the others in the samples in \href{https://next-week-tonight.github.io/NWT//#style_control.single_variate}{S4.1} which enables a more granular manual control over synthesis.
 \end{itemize}

\subsubsection{Video Memcode decoder}
The video Memcode decoder takes previously generated elements (previous frames for FAR and previous Memcodes for MAR) as input. These inputs are embedded and then passed through a series of causal convolution layers with dilations. In the case of FAR, the 3D convolutions are only causal and only use dilations on the time dimension; the space convolutions are bidirectional and do not dilate. 

For gradient stability, we use residual connections every few dilated convolution layers. The conditioning audio and style (outputs of the audio and style encoders) are concatenated to the decoder's activations and linearly projected before each residual block.


In inference, we use the caching strategy from \citet{paine2016fast} for the decoder to speed up generation and reduce the computational complexity of the autoregressive sampling process.

\subsection{Episode embedding} \label{episode_embedding_appendix}
To make our models generalize across episodes of LWT data, we add episode embedding tables in both the dVAE-Adv model and the prior autoregressive model. 

We use the episode index to query from the episode embedding of dVAE-Adv and we concatenate the queried embedding to the hard multi-head attention output. The episode embedding is tiled to match the size of the Memcodes. We use that concatenation as the dVAE-Adv decoder input.

For the prior autoregressive model, we use the episode index to query from the episode embedding and we concatenate the queried embedding to the decoder activations along with the audio and style conditioning. We tile the episode embedding before concatenation. This concatenated vector is passed through the linear projection and used as input for each residual block.

\section{Model training and parameters}\label{hyperparameters_appendix}

\subsection{Training algorithm}\label{appendix_training_alg}

For completeness, we present a pseudocode of the training procedure of dVAE-Adv in Algorithm \ref{training_alg}.

\begin{algorithm}
   \caption{Training procedure}
   \label{training_alg}
\begin{algorithmic}
\State {\bfseries Input:} Initial encoder parameters $\phi$, initial decoder parameters $\theta$ and initial critics parameters $\theta_\mathrm{critics}$. A uniform distribution $\mathcal{U}$, training dataset $\mathbb{P}_r$, the number of training updates $N_\mathrm{iter}$ and Adam optimizer parameters ($\eta$, $\beta_1$, $\beta_2$, $\epsilon$).\\
\For{$\mathrm{iter} \leftarrow 0$ to $N_\mathrm{iter}$}
  \State // Forward pass
  \State $x \sim \mathbb{P}_r$
  \State $z \sim q_\phi(z|x)$
  \State $\mathcal{L}_\mathrm{prior} \leftarrow D_\KL \left(q_{\phi}(z|x) || p(z)\right) $
  \State $\tilde{x} \sim p_\theta(x|z)$
  \State $\mathcal{L}_{\mathrm{gan}} \leftarrow \sum_{c \in \mathbb{S}_C} \left[ - D_c(\tilde{x}) \right]$
  \State $\mathcal{L}_{\mathrm{recon}} \leftarrow \sum_{c \in \mathbb{S}_C} \sum_{l \in \mathbb{S}_{L_{c}}} - \log p_{\theta}(D^{l}_{c}(x)|z)$
  \State $u \sim \mathcal{U}(0, 1)$
  \State $\hat{x} \leftarrow u x + \left(1 - u \right) \tilde{x}$
  \State $\mathcal{L}_{\mathrm{adv}} = - \mathcal{L}_\mathrm{gan} + \sum_{c \in \mathbb{S}_C} \left(- D_c(x) + \lambda \left(\left \| \nabla_{\hat{x}}D_c(\hat{x}) \right \|_2 - 1\right)^2 \right)$\\
  \State // Update parameters
  \State $\theta_\mathrm{critics} \leftarrow \mathrm{Adam}\left(\nabla_{\theta_\mathrm{critics}} \mathcal{L}_{\mathrm{adv}}, \theta_\mathrm{critics}, \eta, \beta_1, \beta_2, \epsilon \right) $ 
  \State $\theta \leftarrow \mathrm{Adam}\left (\nabla_\theta \left( \mathcal{L}_{\mathrm{recon}} + \gamma \mathcal{L}_{\mathrm{gan}} \right), \theta, \eta, \beta_1, \beta_2, \epsilon \right) $ 
  \State $\phi \leftarrow \mathrm{Adam}\left (\nabla_\phi \left( \mathcal{L}_{\mathrm{recon}} + \gamma \mathcal{L}_{\mathrm{gan}} + \beta \mathcal{L}_\mathrm{prior} \right), \phi, \eta, \beta_1, \beta_2, \epsilon \right) $ 

\EndFor
\end{algorithmic}
\end{algorithm}

\subsection{Parameters}

\subsubsection{dVAE-Adv}\label{dvaeadv_parameters_appndex}

We train dVAE-Adv for $150,000$ iterations with a batch size of 16 distributed across 16 NVIDIA A100 40GB GPUs using TensorFloat32. We use Adam optimizer \cite{adamoptimizer} with $\beta_1 = 0.9$, $\beta_2 = 0.999$, $\epsilon = 1e^{-7}$ and a constant learning rate of $1e^{-4}$ for both the generator and the critics. For the loss parameters, we set the gradient penalty weight $\lambda = 10$. Large $\gamma$ values encourage dVAE-Adv to generate realistic videos while sacrificing consistency with the target, and smaller values ensure better matching to the target but are more prone to high frequency averaging problems. In this work, we set $\gamma=5e^{-3}$ which uses the adversarial loss as a good blur regularization, while matching the target's contents. We use the KL-annealing trick to logistically increase $\beta$\footnote{When computing the log likelihood and KL divergence in expectation over the individual critic feature maps and individual Memcodes, we find using $\beta_\mathrm{norm}$ from $\beta$-VAE \cite{Higgins2017betaVAELB} to work well.} from 0 to 1 with a midpoint of $5,000$ steps and $1e^{-3}$ growth rate. The KL-annealing isn't necessary but speeds up convergence. We limit the videos to be $1$ second long ($30$ frames) during the training of dVAE-Adv, and we find the trained model to be able to generate arbitrary length videos during inference without decay in quality.

\begin{table} 
\caption{Generator and the three critics architectures for dVAE-Adv. We denote residual blocks with Res and Nearest Neighbor Upsampling with NNU. We also note Multihead hard attention as MHA. All layers use LeakyReLU with $\alpha=0.2$ as an activation unless otherwise specified.}
\label{dvae_adv_architecture}
\begin{minipage}{.45\textwidth}
\subcaption{Generator Architecture}%
\begin{tabular}{c}
\toprule
$4\times4,\ \mathrm{stride}=1,\ \mathrm{Conv2D},\ 32$ \\
\midrule
$\mathrm{Res}\ 3\times:\ 4\times4,\ \mathrm{stride}=1,\ \mathrm{Conv2D},\ 64$ \\
$4\times4,\ \mathrm{stride}=2,\ \mathrm{Conv2D},\ 64$ \\
\midrule
$\mathrm{Res}\ 4\times:\ 4\times4,\ \mathrm{stride}=1,\ \mathrm{Conv2D},\ 128$ \\
$4\times4,\ \mathrm{stride}=2,\ \mathrm{Conv2D},\ 128$ \\
\midrule
$\mathrm{Res}\ 6\times:\ 4\times4,\ \mathrm{stride}=1,\ \mathrm{Conv2D},\ 256$ \\
$4\times4,\ \mathrm{stride}=2,\ \mathrm{Conv2D},\ 256$ \\
\midrule
$\mathrm{Res}\ 3\times:\ 4\times4,\ \mathrm{stride}=1,\ \mathrm{Conv2D},\ 256$ \\
$4\times4,\ \mathrm{stride=2},\ \mathrm{Conv2D},\ 256$ \\
\midrule
$4\times2\times2,\ \mathrm{stride}=1,\ \mathrm{Conv3D},\ 256$ \\
$2\times2\times2,\ \mathrm{stride}=1,\ \mathrm{Conv3D},\ 512$ \\
$2\times2\times2,\ \mathrm{stride}=1,\ \mathrm{Conv3D},\ 512$ \\
\midrule
$\mathrm{heads}=8,\ \mathrm{codebook}=8$ \\ 
$\mathrm{codebook\_depth}=1024,\ \mathrm{MHA},\ 1024$ \\
\midrule
$2\times2\times2,\ \mathrm{stride}=1,\ \mathrm{Conv3D},\ 512$ \\
$2\times2\times2,\ \mathrm{stride}=1,\ \mathrm{Conv3D},\ 512$ \\
$4\times2\times2,\ \mathrm{stride}=1,\ \mathrm{Conv3D},\ 256$ \\
\midrule
$4\times4,\ \mathrm{stride}=2,\ \mathrm{NNU}\ +\ \mathrm{Conv2D},\ 256$ \\
$\mathrm{Res}\ 3\times:\ 4\times4,\ \mathrm{stride}=1,\ \mathrm{Conv2D},\ 256$ \\
\midrule
$4\times4,\ \mathrm{stride}=2,\ \mathrm{NNU}\ +\ \mathrm{Conv2D},\ 256$ \\
$\mathrm{Res}\ 6\times:\ 4\times4,\ \mathrm{stride}=1,\ \mathrm{Conv2D},\ 256$ \\
\midrule
$4\times4,\ \mathrm{stride}=2,\ \mathrm{NNU}\ +\ \mathrm{Conv2D}\ 128$ \\
$\mathrm{Res}\ 4\times:\ 4\times4,\ \mathrm{stride}=1,\ \mathrm{Conv2D},\ 128$ \\
\midrule
$4\times4,\ \mathrm{stride}=2,\ \mathrm{NNU}\ +\ \mathrm{Conv2D},\ 64$ \\
$\mathrm{Res}\ 3\times:\ 4\times4,\ \mathrm{stride}=1,\ \mathrm{Conv2D},\ 64$ \\
\midrule
$4\times4,\ \mathrm{stride}=1,\ \mathrm{Conv2D},\ \tanh,\ 3$ \\
\bottomrule
\end{tabular}
\end{minipage}\qquad
\begin{minipage}{.45\textwidth}
\subcaption{Critics architectures}
\begin{tabular}{c}
\toprule
Conv3D critic \\
\midrule
$1\times4\times4,\ \mathrm{stride}=1,\ 32$ \\
$2\times4\times4,\ \mathrm{stride}=1\times4\times4,\  64$ \\
$2\times4\times4,\ \mathrm{stride}=1\times4\times4,\  512$ \\
$4\times2\times2,\ \mathrm{stride}=1,\  512$ \\
$1\times1\times1,\ \mathrm{stride}=1,\  \mathrm{Linear},\ 1$ \\
\midrule
First Conv2D critic \\
\midrule
$2\times2,\ \mathrm{stride}=1,\  32$ \\
$4\times4,\ \mathrm{stride}=2\times2,\ \mathrm{groups}=16,\  128$ \\
$4\times4,\ \mathrm{stride}=2\times2,\ \mathrm{groups}=64,\  512$ \\
$4\times4,\ \mathrm{stride}=2\times2,\ \mathrm{groups}=64,\  512$ \\
$2\times2,\ \mathrm{stride}=1\times1,\  512$ \\
$1\times1,\ \mathrm{stride}=1\times1,\  \mathrm{Linear},\ 1$ \\
\midrule
Second Conv2D critic \\
\midrule
$4\times4,\ \mathrm{stride}=1,\   32$ \\
$8\times8,\ \mathrm{stride}=4\times4,\ \mathrm{groups}=16,\   128$ \\
$4\times4,\ \mathrm{stride}=2\times2,\ \mathrm{groups}=64,\   512$ \\
$4\times4,\ \mathrm{stride}=2\times2,\ \mathrm{groups}=256,\   1024$ \\
$4\times4,\ \mathrm{stride}=2\times2,\ \mathrm{groups}=256,\   1024$ \\
$2\times2,\ \mathrm{stride}=1\times1,\   1024$ \\
$1\times1,\ \mathrm{stride}=1\times1,\   \mathrm{Linear},\  1$ \\
\bottomrule
\end{tabular}
\end{minipage}
\end{table}

We show the architecture details of dVAE-Adv in Table \ref{dvae_adv_architecture}. We found it beneficial for perceptual quality to use grouped convolutions with large number of filters in the critics. That allowed our critics to learn a large number of features about the video frames without exploding in number of parameters. 

\subsubsection{Prior autoregressive model}\label{parm_parameters_appendix}

We train our prior autoregressive models for $250,000$ steps with a batch size of 64 distributed across 8 NVIDIA A100 GPUs Using TensorFloat32. We use Adam optimizer with $\beta_1 = 0.9$, $\beta_2 = 0.999$, $\epsilon = 1e^{-7}$ and a learning rate of $1e^{-4}$ exponentially decaying by half every $50,000$ updates until it reaches a minimum of $1e^{-5}$. To minimize the KL divergence between the style posterior $q(z_l|z)$ and the prior $p(z_l)$, we set a constant $\beta_{l}=1$ and we set a minimum style latent posterior std of $\exp(-4)$. We also limit our samples to 32 frames during training and we ensure the autoregressive decoders' receptive fields do not exceed 27 frames, which we found sufficient to learn the task. We do not observe a posterior collapse despite the strong autoregressive decoder \cite{bowman2016generating, yang2017improved, kim2018semiamortized}, likely due to the complexity of video data distribution compared to speech. During inference, the models are capable of generating arbitrary length videos without any decay in quality compared to training. 

In Tables \ref{far_architecture_details} and \ref{mar_architecture_details}, we present the architecture details of our prior autoregressive models. When training FAR models, we found the trick of downsampling Memcodes spatially before convolving them on time to save memory and allow for training larger models without hurting performance. Since MAR models have more autoregressive steps, they inherently use more memory during training. They also cannot make use of the spatial downsampling trick, thus our MAR models had to be much smaller in number of parameters compared to FAR.

Since the latent shape is not square ($16\times14$), we use padding to make it $16\times16$ for the FAR model, which makes parameters such as kernel shapes and strides easier to pick. Later, we simply drop the padded regions from the output Memcodes before decoding them to the pixel domain.

\begin{table} 
\caption{FAR model architecture. we note Nearest Neighbor Upsampling with NNU. We also note Spatially Connected 3D layers with SC3D and we represent layers whose inputs are concatenated to audio encoder and style encoder outputs with Concat. All layers use the Mish activation function unless otherwise specified.}
\label{far_architecture_details}
\begin{center}
\begin{tabular}{c  c}
\toprule
\multicolumn{2}{c}{Audio Encoder} \\
\midrule
Conv1D & $3,\ \mathrm{stride}=3,\ 256 $ \\
Conv1D & $3,\ \mathrm{stride}=1,\ 256 $ \\
Conv1D & $3,\ \mathrm{stride}=1,\ 256 $ \\
Conv1D & $3,\ \mathrm{stride}=1,\ 512 $ \\
Conv1D & $3,\ \mathrm{stride}=1,\ \mathrm{dilation}=2,\ 512 $ \\
Conv1D & $3,\ \mathrm{stride}=1,\ \mathrm{dilation}=4,\ 512 $ \\
Conv1D & $3,\ \mathrm{stride}=1,\ \mathrm{dilation}=8,\ 1024 $ \\
NNU + SC3D & $3\times 2\times 2,\ \mathrm{stride}=1\times 2 \times 2,\ 256$ \\
SC3D & $2\times: 3 \times 2 \times 2,\ \mathrm{stride}=1,\ 256$ \\
NNU + SC3D & $3\times 2\times 2,\ \mathrm{stride}=1\times 2\times 2,\ 128$ \\
SC3D & $2\times: 3\times 2\times 2,\ \mathrm{stride}=1,\ 128$ \\
\midrule
\multicolumn{2}{c}{Style Encoder} \\
\midrule
Embedding & $\mathrm{vocab\_size}=8,\ \mathrm{depth}=128$ \\
Conv3D & $2 \times 2\times 2,\ \mathrm{stride}=1\ \times 2\times 2,\  256$ \\
Conv3D & $2 \times 2 \times 2,\ \mathrm{stride}=1,\  512$ \\
Conv3D & $3\times2\times2,\ \mathrm{stride}=1,\ \mathrm{dilation}=3\times1\times1,\  512 $ \\
Conv3D & $3\times2\times2,\ \mathrm{stride}=1,\ \mathrm{dilation}=9\times1\times1,\  512 $ \\
Global average pooling 3D & $-- $ \\
Posterior Conv3D & $1\times1\times1,\ \mathrm{stride}=1,\  16\times2 $ \\
NNU + SC3D & $3\times2\times2,\ \mathrm{stride}=1\times2\times2,\   256 $ \\
SC3D & $2\times: 3\times2\times2,\ \mathrm{stride}=1,\   256 $ \\
NNU + SC3D & $3\times2\times2,\ \mathrm{stride}=1\times2\times2,\   128 $ \\
SC3D & $2\times: 3\times2\times2,\ \mathrm{stride}=1,\   128 $ \\
\midrule
\multicolumn{2}{c}{Audio Decoder} \\
\midrule
Embedding & $ \mathrm{vocab\_size}=8,\ \mathrm{depth}=128 $ \\
Conv2D & $ 4\times4,\ \mathrm{stride}=2\times2,\   1024 $ \\
Conv2D & $ 4\times4,\ \mathrm{stride}=1\times1,\   1024 $ \\
Conv2D & $ 4\times4,\ \mathrm{stride}=2\times2,\   1024 $ \\
Conv3D & $ \mathrm{Concat}\  1\times1\times1,\ \mathrm{stride}=1,\  \mathrm{Linear},\ 1024 $ \\
Res block1 & $ 3\times2\times2,\ \mathrm{dilation}=1\times1\times1,\    1024 $ \\ 
Res block1 & $ 3\times2\times2,\ \mathrm{dilation}=1\times1\times1,\    4096 $ \\
Res block1 & $ 3\times1\times1,\ \mathrm{dilation}=2\times1\times1,\    1024 $ \\
Conv3D & $ \mathrm{Concat}\  1\times1\times1,\ \mathrm{stride}=1,\   \mathrm{Linear},\ 1024 $ \\
Res block2 & $ 3\times2\times2,\ \mathrm{dilation}=1\times1\times1,\    1024 $ \\ 
Res block2 & $ 3\times1\times1,\ \mathrm{dilation}=2\times1\times1,\    1024 $ \\
Conv3D & $ \mathrm{Concat}\  1\times1\times1,\ \mathrm{stride}=1,\  \mathrm{Linear},\  1024 $ \\
Res block3 & $ 3\times2\times2,\ \mathrm{dilation}=1\times1\times1,\    1024 $ \\ 
Res block3 & $ 3\times2\times2,\ \mathrm{dilation}=2\times1\times1,\    4096 $ \\
Res block3 & $ 3\times1\times1,\ \mathrm{dilation}=4\times1\times1,\    1024 $ \\
NNU + Conv2D & $ 4\times4,\ \mathrm{stride}=2\times2,\  1024 $ \\
Conv2D & $ 4\times4,\ \mathrm{stride}=1\times1,\  1024\  $ \\
NNU + Conv2D & $ 4\times4,\ \mathrm{stride}=2\times2,\  1024 $ \\
Conv3D & $ 1\times1\times1,\ \mathrm{stride}=1\ \mathrm{Linear},\  8\times8 $ \\
\bottomrule
\end{tabular}
\end{center}
\end{table}

\begin{table} 
\caption{MAR model architecture. we note Nearest Neighbor Upsampling with NNU. We also note Spatially Connected 3D layers with SC3D and we represent layers whose inputs are concatenated to audio encoder and style encoder outputs with Concat. All layers use the Mish activation function unless otherwise specified.}
\label{mar_architecture_details}
\begin{center}
\begin{tabular}{c  c}
\toprule
\multicolumn{2}{c}{Audio Encoder} \\
\midrule
Conv1D & $3,\ \mathrm{stride}=3,\  256$ \\
Conv1D & $3,\ \mathrm{stride}=1,\  256$ \\
Conv1D & $3,\ \mathrm{stride}=1,\  256$ \\
Conv1D & $3,\ \mathrm{stride}=1,\  512$ \\
Conv1D & $3,\ \mathrm{stride}=1,\ \mathrm{dilation}=2,\ 512$ \\
Conv1D & $3,\ \mathrm{stride}=1,\ \mathrm{dilation}=4,\ 512$ \\
Conv1D & $3,\ \mathrm{stride}=1,\ \mathrm{dilation}=8,\ 1024$ \\
NNU + SC3D & $3\times2\times2, \mathrm{stride}=1\times2\times2,\ 256$ \\
SC3D & 2$\times: 3\times2\times2, \mathrm{stride}=1,\ 256$ \\
NNU + SC3D & $3\times2\times2, \mathrm{stride}=1\times2\times2,\ 128$ \\
SC3D & 2$\times: 3\times2\times2, \mathrm{stride}=1,\ 128$ \\
NNU + SC3D & $3\times2\times2, \mathrm{stride}=1\times2\times2,\ 64$ \\
SC3D & 2$\times: 3\times2\times2, \mathrm{stride}=1,\ 64$ \\
NNU + SC3D & $3\times2\times2, \mathrm{stride}=1\times2\times2,\ 64$ \\
SC3D & 2$\times: 3\times2\times2, \mathrm{stride}=1,\ 64$ \\
\midrule
\multicolumn{2}{c}{Style Encoder} \\
\midrule
Conv3D & $2\times2\times2,\ \mathrm{stride}=1\times2\times2,\ 256$ \\
Conv3D & $2\times2\times2,\ \mathrm{stride}=1,\  512$ \\
Conv3D & $3\times2\times2,\ \mathrm{stride}=1,\ \mathrm{dilation}=3\times1\times1,\  512$ \\
Conv3D & $3\times2\times2,\ \mathrm{stride}=1,\ \mathrm{dilation}=9\times1\times1,\  512$ \\
Global average pooling 3D & -- \\
Posterior Conv3D & $1\times1\times1,\ \mathrm{stride}=1,\ 16\times2$ \\
NNU + SC3D & $3\times2\times2,\ \mathrm{stride}=1\times2\times2,\  256$ \\
SC3D & $2\times: 3\times2\times2,\ \mathrm{stride}=1,\  256$ \\
NNU + SC3D & $3\times2\times2,\ \mathrm{stride}=1\times2\times2,\ 128$ \\
SC3D & $2\times: 3\times2\times2,\ \mathrm{stride}=1,\ 128$ \\
NNU + SC3D & $3\times2\times2,\ \mathrm{stride}=1\times2\times2,\ 64$ \\
SC3D & $2\times: 3\times2\times2,\ \mathrm{stride}=1,\ 64$ \\
NNU + SC3D & $3\times2\times2,\ \mathrm{stride}=1\times2\times2,\ 64$ \\
SC3D & $2\times: 3\times2\times2,\ \mathrm{stride}=1,\ 64$ \\
\midrule
\multicolumn{2}{c}{Audio Decoder} \\
\midrule
Embedding & $ \mathrm{vocab\_size}=8,\ \mathrm{depth}=64 $ \\
Conv1D & $\mathrm{concat}\ 1,\ \mathrm{stride}=1,\ \mathrm{Linear},\ 512$ \\
Res block1 & $4,\ \mathrm{dilation}=1,\ \mathrm{Conv1D},\ 1024$ \\ 
Res block1 & $4,\ \mathrm{dilation}=2,\ \mathrm{Conv1D},\ 2048$ \\
Res block1 & $4,\ \mathrm{dilation}=4,\ \mathrm{Conv1D},\ 1024$ \\
Conv1D & $\mathrm{concat},\  1,\ \mathrm{stride}=1,\ \mathrm{Linear},\ 512$ \\
Res block2 & $4,\ \mathrm{dilation}=8,\ \mathrm{Conv1D},\ 1024$ \\ 
Res block2 & $4,\ \mathrm{dilation}=16,\ \mathrm{Conv1D},\ 2048$ \\ 
Res block2 & $4,\ \mathrm{dilation}=32,\ \mathrm{Conv1D},\ 1024$ \\
Conv1D & $\mathrm{concat}\  1,\ \mathrm{stride}=1\ \mathrm{Linear},\ 512$ \\
Res block3 & $4,\ \mathrm{dilation}=64,\ \mathrm{Conv1D},\ 1024$ \\ 
Res block3 & $4,\ \mathrm{dilation}=128,\ \mathrm{Conv1D},\ 1024$ \\
Res block3 & $4,\ \mathrm{dilation}=256,\ \mathrm{Conv1D},\ 1024$ \\
Res block3 & $4,\ \mathrm{dilation}=512,\ \mathrm{Conv1D},\ 1024$ \\
Conv1D & $1,\ \mathrm{stride}=1,\ \mathrm{Linear},\ 8\times8$ \\
\bottomrule
\end{tabular}
\end{center}
\end{table}

\subsection{FAR vs MAR} \label{far_vs_mar}

In Table \ref{model_speeds}, we present the FAR and MAR model sizes alongside their training and inference speeds. Despite the fact that MAR models are more stable in inference and yield better MOS scores on average, FAR models are much faster. We also find that FAR models tend to be more emotive, likely due to the bigger number of parameters, and cover a larger range of body movements compared to MAR. We find it useful to have both models for different use cases; where FAR is usable in real time inference, MAR would be a better choice on large scale slow video inference.

\section{Data preparation}\label{dataprep_details_appendix}

We prepared our dataset by the following procedure:

\begin{enumerate}

\item download videos using \texttt{youtube-dl} given a list of $219$ video URLs, all sourced from the official Last Week Tonight channel on YouTube: \url{https://www.youtube.com/user/LastWeekTonight}.

\item split episodes using H.264 codec key frames as boundaries, which divides videos into segments of a maximum of $8.37$ seconds using \texttt{ffmpeg}:

\lstset{language=bash}
\begin{lstlisting}
$ ffmpeg -y -i $INPUT_FILEPATH \
-acodec copy -f segment -segment_time 4 \ 
-vcodec copy -reset_timestamps 1 -map 0 \ 
$OUTPUT_FILEPATH
\end{lstlisting}

\item crop videos from $507$ to $1130$ left to right, and $0$ to $712$ bottom to top.

\item for every clip greater than $2$ seconds, trim $0.25$ seconds from the beginning and end, resulting in a maximum clip duration of $7.87$ seconds

\item manually review to remove clips where any of the following occur
    \begin{itemize}
        \item subject is not in frame
        \item subject is holding objects
        \item there are person(s) other than subject visible
        \item there are intruding overlays inserted in post production
        \item there is a title screen transition
        \item subject is not sitting at desk, e.g. standing or conducting an interview
    \end{itemize}

\item To protect our models from trying to learn the seemingly random side overlays in the videos, we apply an algorithmic heuristic to mask them using the following OpenCV \cite{opencv_library} functions. The following is applied to each frame in every video of the dataset:
\begin{enumerate}
    \item convert frame to grayscale \begin{verbatim}gray_image = cv2.cvtColor(color_image, cv2.COLOR_BGR2GRAY)\end{verbatim}
    \item binarize the frame \begin{verbatim}binarized_image = cv2.threshold(gray_image, 185, 255, 0)[1]\end{verbatim}
    \item find contours \begin{verbatim}contours = cv2.findContours(binarized_image, cv2.RETR_EXTERNAL, 
    cv2.CHAIN_APPROX_SIMPLE)[0]\end{verbatim}
    \item for each contour, find the bounding rectangle \begin{verbatim}x, y, w, h = cv2.boundingRect(contour)\end{verbatim}
    \item for each bounding rectangle, if the area is greater than 4000 pixels, smaller than 100000 pixels, has a height greater than 80 pixels, and the top left corner of the bounding box either within 20 pixels from the left hand side of the frame, or starts more than 550 pixels from the top of the frame, then this bounding box is used to overlay the original frame with a white rectangle.
\end{enumerate}

\item using \texttt{moviepy}, scale clips to $256 \times 224$ and interpolate to $30$ frames per second (FPS), preserving duration (downloaded data contains a mix of frame rates including $\frac{30000}{1001}$ and $30$ FPS).

\item extract audio from video clips, interpolating to a sample rate of $24000$, convert to mono and generate Mel spectrograms using the parameters in Table \ref{mel_params}.

\item Scale video pixels to a range of $[-1, 1]$ with min-max scaling: $x = \frac{x}{127.5} - 1$

\begin{table}
\caption{Mel spectrogram parameters.}
\label{mel_params}
\begin{center}
\begin{tabular}{ll}
\toprule
parameter & value   \\
\midrule
n\_mels & $80$ \\
n\_fft & $1024$ \\
hop\_length & $200$ \\
win\_length & $800$ \\
center & True \\
pad\_mode & \texttt{constant} \\
power & $2.$ \\
fmin & $0.$ \\
fmax & $8000.$ \\
\bottomrule
\end{tabular}
\end{center}
\end{table}

\item split into train, validation, and test sets stratified by episodes with ratios of approximately $90\%$, $5\%$, and $5\%$ respectively.

\end{enumerate}

\newpage 

\section{Experimental setup of MOS} \label{mos_appendix}

We randomly selected $1$ clip from $50$ randomly selected episodes from the test set to evaluate in all three experiments. The ground truth and synthesized videos (by our models and the baselines used for comparison) are sent to Amazon's Mechanical Turk where each sample is rated by at least 30 participants initially, on a scale from 1 to 5 with 1 point increments. 

The GUI for the overall naturalness MOS experiment is shown in Figure \ref{mturk_gui_1}. The participants are required to complete 4 sections: qualification, setup, training, and rating. The qualification section requires the participants to answer English for mother tongue, to state that they are using in-ear headphones or over-the-ear headphones, and that they have normal hearing ability. If they do not provide these answers, the participant will be notified that they do not qualify. The setup section requires each participant to watch $4$ pairs of altered ground truth videos and choose which one is of higher quality. Three of the pairs have an audio shift, a video with a smaller audio shift is considered of higher quality. One of the pairs has visual noise. Participants must answer at least 3 out of the 4 comparison questions correctly to proceed. This ensures they are able pay attention to both video and audio details. The training section includes the subjective rating task. There are 6 videos that are used in this section. The ratings from the training section are used for calibration and are not included in the participant's scores. The rating section is similar to the training section, except the submissions here are used to calculate MOS metrics. 

Adapting the guidelines described in CrowdMOS \cite{crowdmos}, the results from the experiments are cleaned via the following post processing procedure to filter out low quality or malicious submissions. Each rating task (HIT) includes 12 videos: 10 to be used for the MOS metric calculation, 1 ``trapping'' clip and 1 ``gold standard'' clip. Trapping clips show a ground truth or synthesized video, but the audio is of a different speaker, clearly instructing the participant not to evaluate the video, and instead give a specific score. We discard all submissions by any participant who fails a trap question at any point. A gold standard clip is a hand selected ground truth video that is deemed very natural and expressive by the authors, and should clearly be given an MOS value of 5. If a participant gives a gold standard video a rating of 3 or below, all of their ratings are discarded. Each of the remaining participants' MOS results are calculated and compared to the global MOS from the remaining population with a Pearson correlation coefficient $r$. Participants whose ratings had a correlation coefficient $r < 0$ had their submissions discarded.

Participants were paid a base value of \$0.35/HIT. If they did 10 or more HITs, they were paid a total of \$0.40/HIT, and if they did more than 10 HITs and had an $r$ value in the top 20\%, they were paid a total of \$0.60/HIT. At an estimated time of 3 minutes per hit on average, the participants were paid \$7-12/hr.

\begin{figure}
\begin{subfigure}{1\textwidth}
    \includegraphics[width=\linewidth]{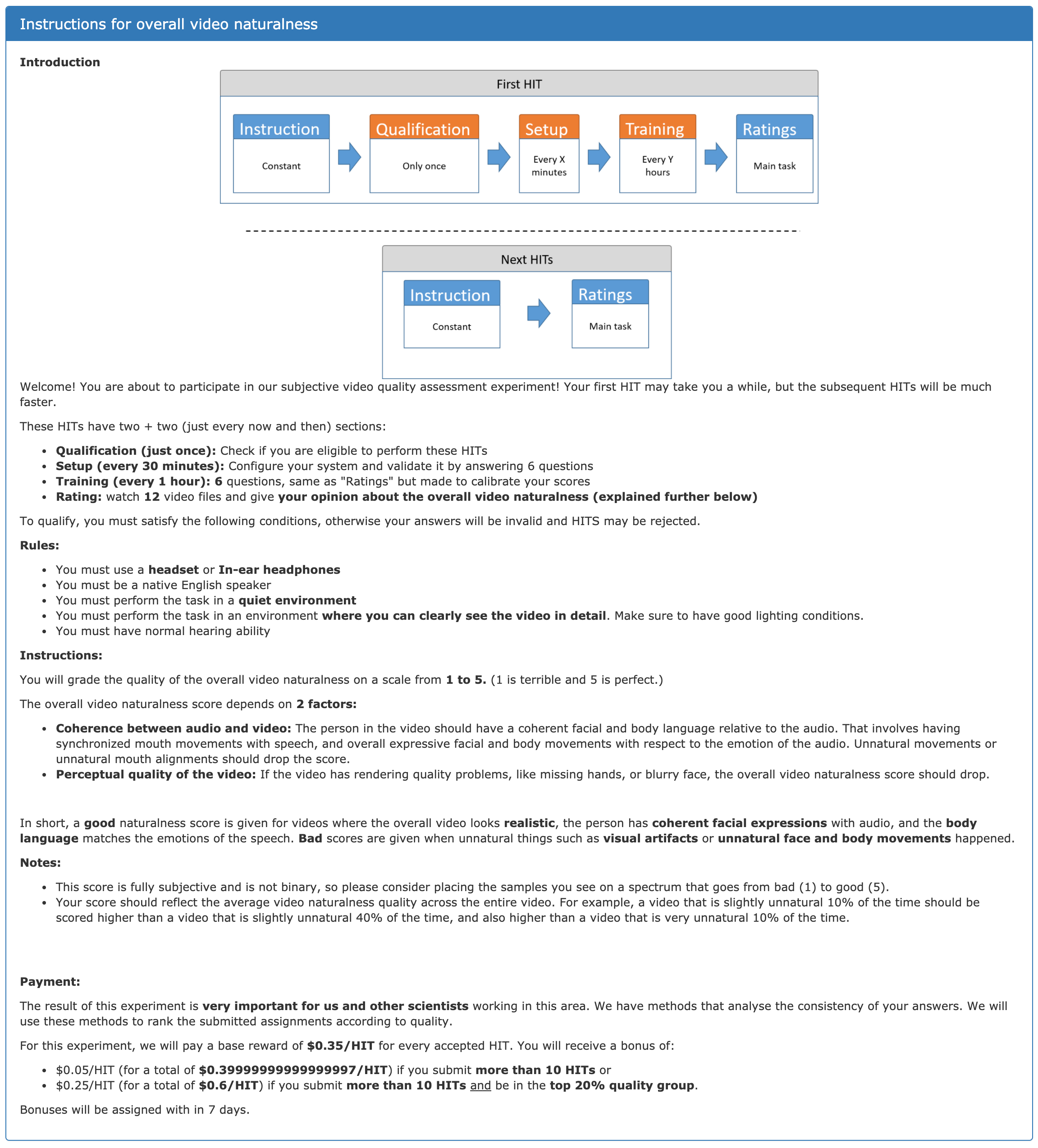}
    \subcaption{Instructions}
    \label{mturk_gui_1}
\end{subfigure}
\end{figure}

\begin{figure}\ContinuedFloat
\begin{subfigure}{1\textwidth}
    \includegraphics[width=\linewidth]{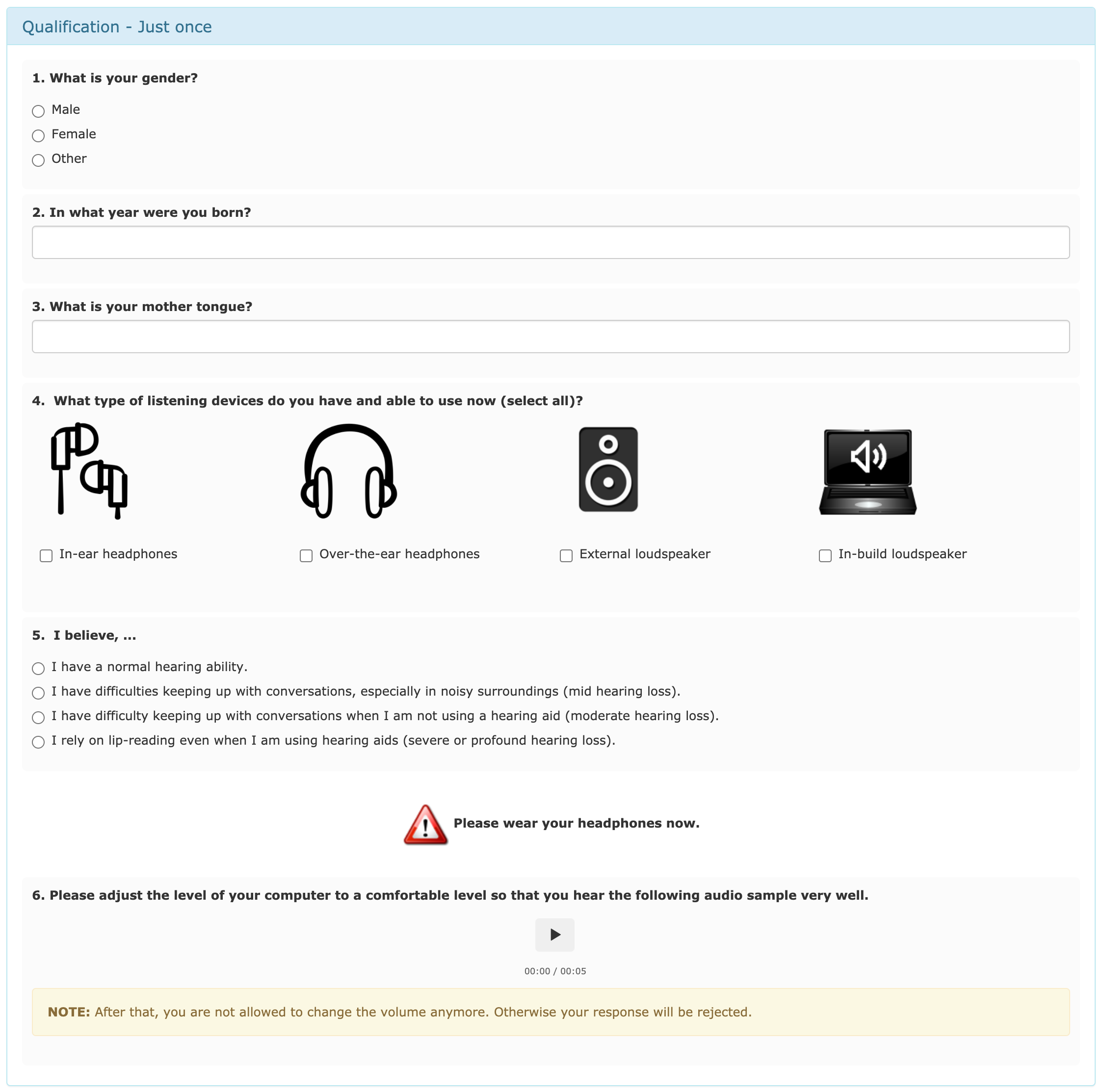}
    \subcaption{Qualification}
\end{subfigure}
\end{figure}

\begin{figure}\ContinuedFloat
\begin{subfigure}{1\textwidth}
    \includegraphics[width=\linewidth]{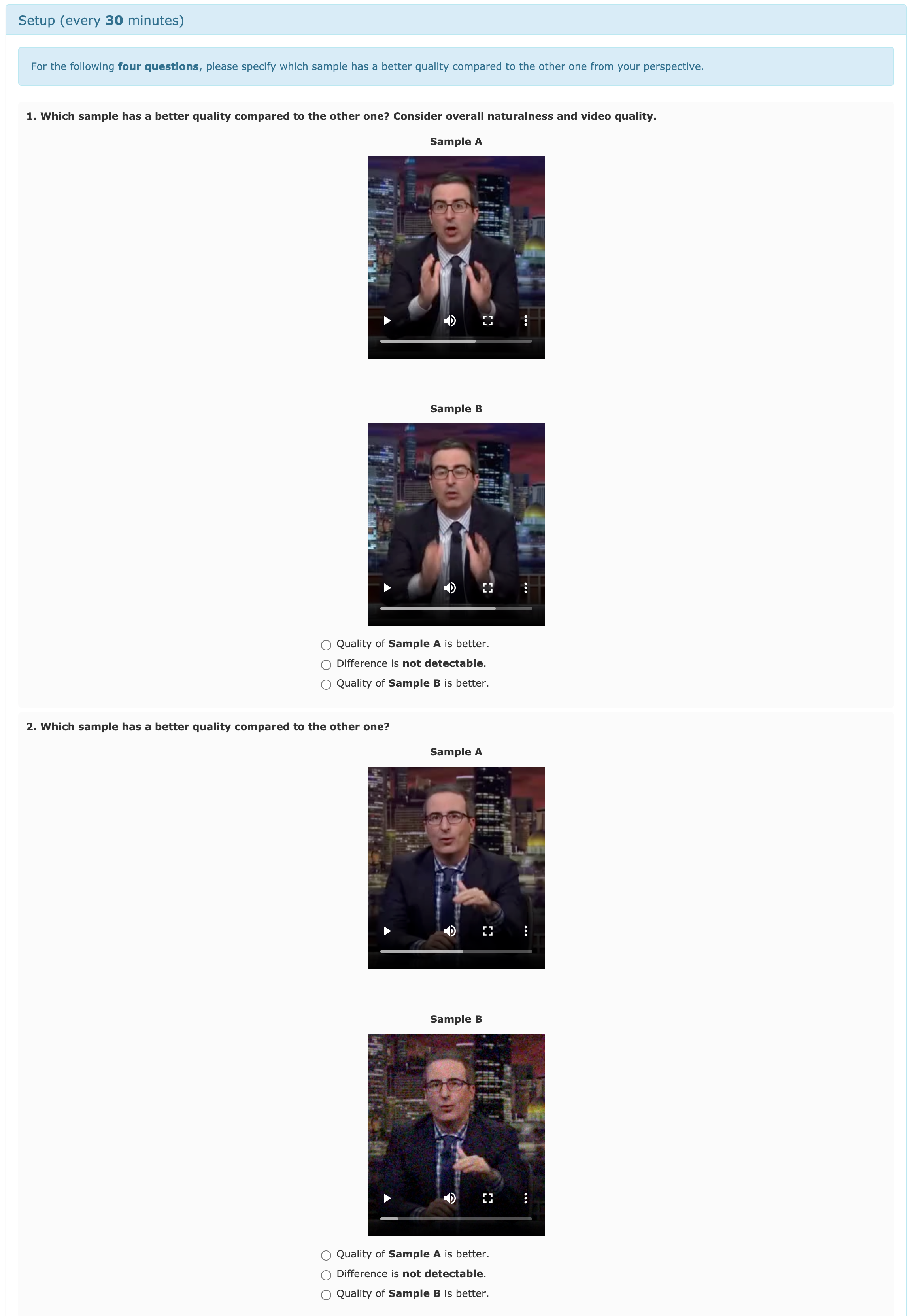}
    \subcaption{Example of comparison questions}
\end{subfigure}
\end{figure}

\begin{figure}\ContinuedFloat
\begin{subfigure}{1\textwidth}
    \includegraphics[width=\linewidth]{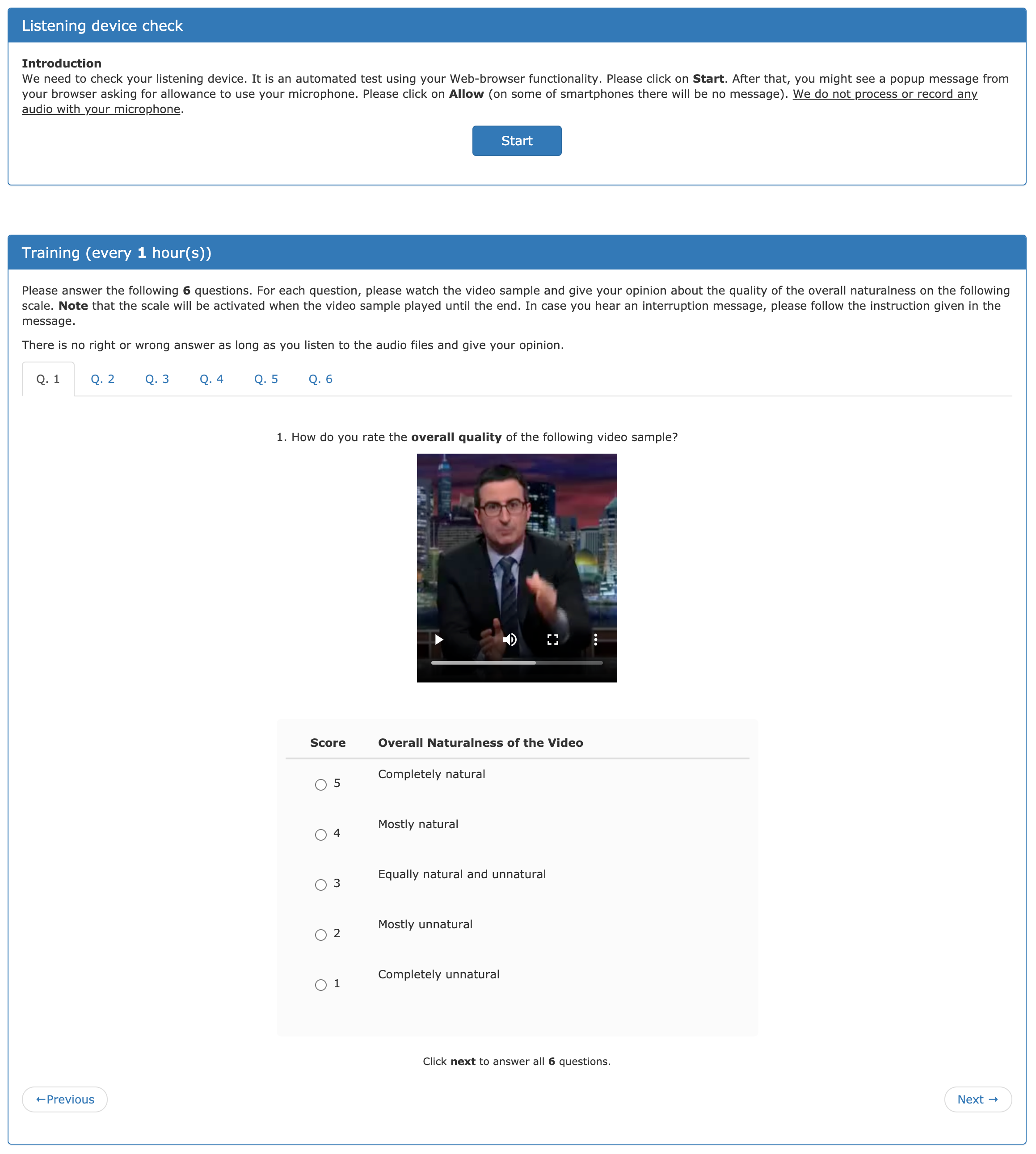}
    \subcaption{Periodic training}
\end{subfigure}
\end{figure}

\begin{figure}\ContinuedFloat
\begin{subfigure}{1\textwidth}
    \includegraphics[width=\linewidth]{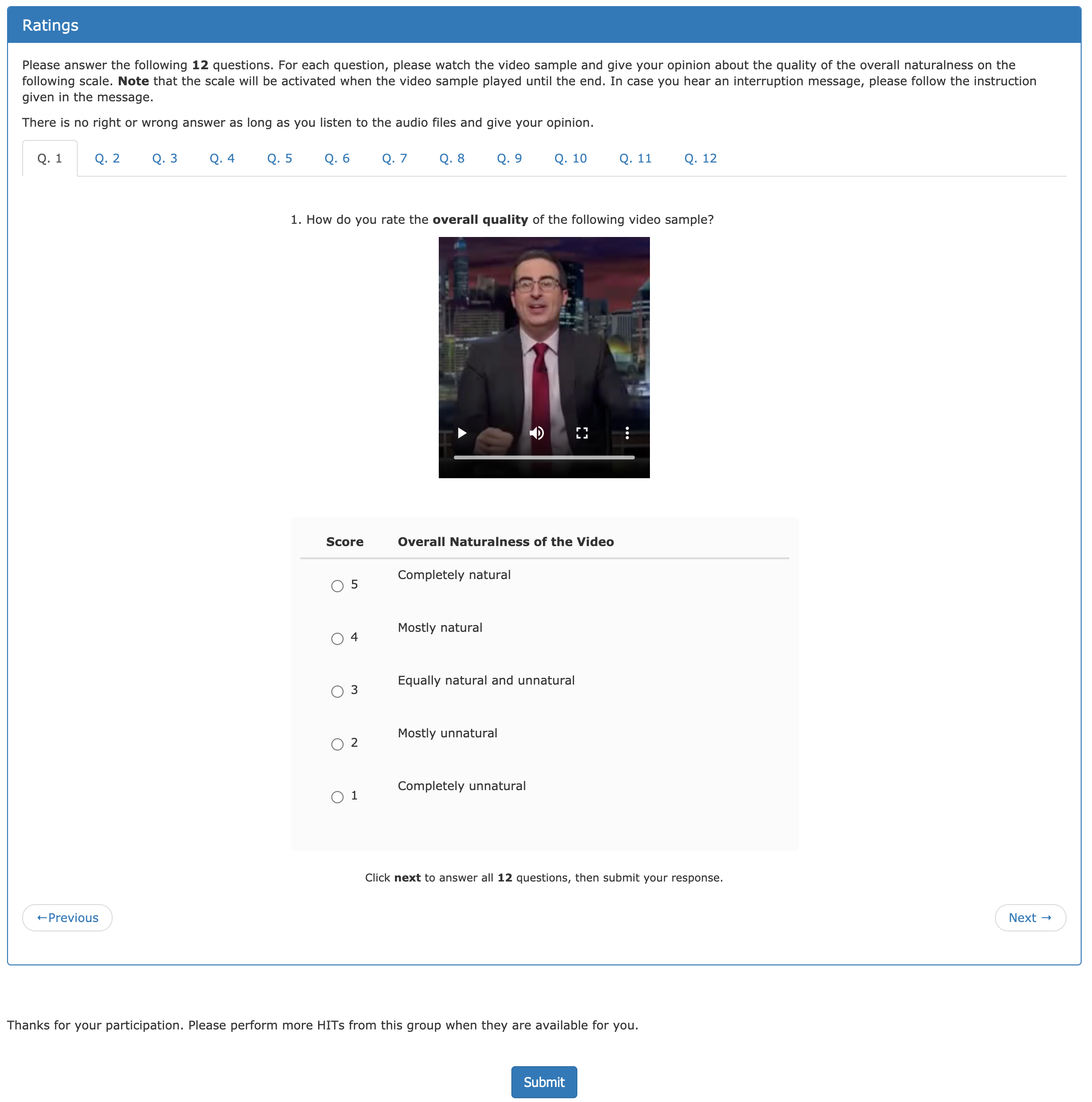}
    \subcaption{Ratings page example}
    \label{mturk_gui_5}
\end{subfigure}
\caption{Mechanical Turk GUI screenshots}
\end{figure}


\end{document}